\documentclass[journal, twocolumn]{IEEEtran}
\usepackage{graphicx}
\graphicspath{{../eps/}}
\DeclareGraphicsExtensions{.eps}

\usepackage{stfloats}
\usepackage{amsmath}
\allowdisplaybreaks[1]
\usepackage{amsfonts,amssymb}
\usepackage{amssymb}
\usepackage{wrapfig}
\usepackage{psfrag}
\usepackage{epstopdf}
\usepackage{cite}
\usepackage{graphicx}
\usepackage{threeparttable}
\usepackage{cases}
\usepackage{subeqnarray}
\usepackage{subcaption}
\usepackage{color}
\usepackage{underscore}
\usepackage{verbatim}
\usepackage{bm}
\usepackage{stfloats}
\usepackage{algorithm}
\usepackage{algorithmicx}
\usepackage{algpseudocode}
\usepackage{microtype}

\makeatletter
\renewcommand{\maketag@@@}[1]{\hbox{\m@th\normalsize\normalfont#1}}%
\makeatother

\begin{document}
	
	\title{You May Use the Same Channel Knowledge Map for Environment-Aware NLoS Sensing and Communication}
	
\author{Di Wu, Zhuoyin Dai, and Yong Zeng,~\IEEEmembership{Fellow, IEEE} \\
	\thanks{D. Wu, Z. Dai and Y. Zeng are with the National Mobile Communications Research Laboratory, Southeast University, Nanjing 210096, China. They are also with Purple Mountain Laboratories, Nanjing 211111, China (e-mail: \{studywudi, zhuoyin\_dai, yong\_zeng\}@seu.edu.cn). (\emph{Corresponding author: Yong Zeng.})}
	}
	
	\maketitle

\begin{abstract}
As one of the key usage scenarios for the sixth generation (6G) wireless networks, integrated sensing and communication (ISAC) provides an efficient framework to achieve simultaneous wireless sensing and communication. However, traditional wireless sensing techniques mainly rely on line-of-sight (LoS) assumptions, i.e., the sensing targets are directly visible to both the sensing transmitter and receiver. This prevents ISAC systems from being applied in complex environments such as the urban low-altitude airspace, which usually suffers from signal blockage and non-line-of-sight (NLoS) multipath propagation. To address this challenge, in this paper, we propose a novel approach to enable environment-aware NLoS ISAC by leveraging the new technique called channel knowledge map (CKM), which was originally proposed for environment-aware wireless communications. One major novelty of our proposed method is that the same CKM built for wireless communication can be directly used to enable NLoS wireless sensing, thus enjoying the benefits of ``killing two birds with one stone''. To this end, the sensing targets are treated as virtual user equipment (UE), and the wireless communication channel priors are transformed into the sensing channel priors, allowing one single CKM to serve dual purposes.
We illustrate our proposed framework using a specific CKM called \emph{channel angle-delay map} (CADM). Specifically, the proposed framework utilizes CADM to derive angle-delay priors of the sensing channel by exploiting the relationship between communication and sensing angle-delay distributions, enabling sensing target localization in the challenging NLoS environment. Extensive simulation results demonstrate significant performance improvements over classic geometry-based sensing methods, which are further validated by Cramér-Rao Lower Bound (CRLB) analysis.
\end{abstract}

\begin{IEEEkeywords}
	Environment-aware integrated communication and sensing (ISAC), NLoS sensing, channel knowledge map (CKM).
\end{IEEEkeywords}

\IEEEpeerreviewmaketitle

\section{Introduction}

Sixth-generation (6G) wireless networks are poised to push the boundaries of wireless connectivity beyond the capabilities of fifth-generation (5G), targeting transformative performance metrics such as Terabit-per-second peak data rates, sub-millisecond latency, ultra-high reliability, and ubiquitous sensing capabilities \cite{you2021towards,zhang20196g}. Driven by emerging applications like autonomous vehicles, smart cities and industrial Internet of Things (IoT), 6G aims to integrate advanced functionalities into a unified wireless framework, enabling seamless interaction between digital and physical worlds \cite{cui2021integrating,wang2024digital}. In contrast to 5G, which primarily focused on enhanced mobile broadband, ultra-reliable low-latency communications, and massive machine-type communications, 6G is expected to incorporate localization and sensing as an additional network service, providing real-time environmental awareness alongside communication \cite{zhang2022enabling,xiao2022overview}. This convergence has given rise to integrated sensing and communication (ISAC), a paradigm that unifies sensing and communication functionalities within the same system, leveraging shared wireless spectrum, signal processing modules, hardware, and network infrastructure \cite{demirhan2023integrated,dai2025tutorial}. ISAC also enables mutual enhancement between wireless communication and sensing, such as communication-assisted sensing for precise localization and sensing-aided communication for improved channel estimation \cite{du2025toward,dong2025communication}, particularly in millimeter-wave (mmWave) massive multiple-input multiple-output (MIMO) systems. With their ultra-wide bandwidths and high spatial resolution, mmWave massive MIMO systems are ideally suited for both high-rate data transmission and precise environmental sensing \cite{heath2016overview,alkhateeb2014channel}.

Significant progress has been made in ISAC systems, particularly in line-of-sight (LoS) scenarios, where the sensing targets are directly visible to both the transmitter and receiver. Various ISAC techniques have been developed, like joint waveform design, beamforming optimization, and interference management to balance radar-like sensing with data transmission, achieving high localization accuracy and spectral efficiency in LoS conditions \cite{xiao2022waveform,wang2024cramer,liu2022joint,zhang2022low,zubow2020deeptxfinder}. For instance, methods such as time-of-arrival (ToA) estimation and angle-based localization leverage direct LoS paths to estimate target locations \cite{xiao2022overview,sturm2011waveform}. However, in complex environments such as urban low-altitude airspace, LoS paths are frequently blocked. As a result, sensing in such environments can only rely on pure non-line-of-sight (NLoS) signal paths, posing a much greater challenge to map the estimated sensing parameters to the location of the sensing targets as more unknowns are involved \cite{richards2005fundamentals,sun2024performance}. Some techniques have been proposed to achieve NLoS localization, including fingerprinting \cite{wang2016csi}, single-base-station (BS) localization exploiting near-field scatterers \cite{zhou2024single}, and reconfigurable intelligent surface (RIS)-aided methods \cite{liu2023ris}. However, they cannot be applied to sensing, because unlike localization tasks, the sensing targets are typically non-cooperative in the sense that they only passively reflect the signals rather than transmitting or receiving them. Specifically, fingerprinting-based localization works by building a database that maps user equipment (UE) location to its wireless communication channel signature such as received signal strength (RSS). However, this approach is unsuitable for sensing. The fundamental reason is that a fingerprint database characterizes the direct channel between the UE and the anchor nodes, whereas sensing relies on the echoes from the passive sensing target. Therefore, the fingerprint database built for localization cannot be used for wireless sensing. Instead, sensing would require its own ``sensing database" that fingerprints target-specific echo characteristics. Creating such databases is quite challenging, if not impossible, as different types of targets (e.g., vehicles, pedestrians) would need a distinct and comprehensive map of their potential echo signatures.
On the other hand, single-BS near-field localization studied in \cite{zhou2024single} first senses near-field scatterers to acquire environment priors, and then leverages the location of near-field scatterers to localize UEs without relying on a LoS link. However, it relies on near-field assumptions, limiting applicability in far-field ISAC. In addition, while RIS-aided methods create virtual LoS relationships for localization by deploying reflective surfaces, they require additional costly and extensive hardware deployments. In fact, rigorous theoretical studies have proven that, without environment or NLoS prior knowledge, NLoS paths cannot provide beneficial information for localization \cite{shen2010fundamental}. This demonstrates the imperative need to develop an innovative framework that can leverage environment prior knowledge to tackle the challenging problem of NLoS sensing for ISAC systems.

Channel knowledge map (CKM) is a promising technique to achieve the above goal \cite{zeng2021toward,zeng2024tutorial}. CKM is a site-specific database capturing location-specific channel characteristics—such as angles of arrival (AoAs), angles of departure (AoDs), delays, and path gains—learned from historical UE-BS interactions. 
This definition distinguishes CKM from the conventional radio map. While radio maps are invaluable for many applications, such as fingerprinting-based localization or interference management \cite{huang2019online}, they usually store signal quality metrics (such as RSS) that are dependent on transmitter and receiver setup and activity. In contrast, CKM aims to directly reflect the intrinsic wireless channel properties of the local environment, independent of transmitter/receiver activity. As ``knowledge" can take various forms, the term CKM aims to signify the much richer, multi-dimensional characteristics of the channel's underlying physical parameters (e.g., AoAs, AoDs, and delays), which are essential for advanced applications like environment-aware communication and sensing.
Recently, extensive research efforts have been devoted to CKM construction and CKM-based environment-aware communication and sensing. For example, to enable AI-based intelligent CKM construction, datasets like CKMImageNet have been developed \cite{wu2025ckmimagenet}. Besides, theoretical analyses have quantified the data requirements for CKM construction \cite{xu2024how}, and various methods have been proposed for CKM construction, leveraging techniques such as generative AI models for sparsely observed data, path matching, and the fusion of environment information like point clouds \cite{jin2024i2i,fu2025ckmdiff,li2023millimeter,wang2024channel, liu2023uav}. 
In terms of CKM utilization for wireless communications, various problems have been investigated, including optimizing resource allocation \cite{yue2024channel,liu2025channel}, improving interference management \cite{jiang2025interference}, enabling efficient beam alignment in mmWave systems \cite{wu2023environment, dai2024prototyping}, facilitating RIS-aided communications \cite{taghavi2023environment}, and enhancing channel prediction \cite{wang2024channel2}.
Moreover, prototyping systems have been developed that have validated its ability to achieve environment-aware mmWave beam alignment, improving connectivity in complex indoor settings \cite{dai2024prototyping}. 
Recent efforts have also explored CKM for sensing applications, such as environment-aware channel estimation with dynamic sensing data \cite{wu2024environment2}, or to enhance CKM construction and applications by sensing the target locations \cite{zhang2025prototyping}. Furthermore, CKM is utilized to enhance sensing by suppressing environmental clutter, improving target parameter estimation \cite{xu2024channel}. These advancements have established a strong foundation for CKM-based sensing. However, they have primarily focused on leveraging CKM for sensing-assisted communication or sensing scenarios by assuming the existence of LoS links, leaving the full potential of exploiting NLoS paths in complex environments underexplored. This gap inspires the development of a novel ISAC framework that directly utilizes NLoS paths to extract target information, leveraging the environmental priors embedded in CKM to address NLoS challenges effectively.

Therefore, in this paper, we try to answer the following question: Can we use the same CKM to achieve both environment-aware NLoS sensing and communication? We show that the answer is affirmative, by proposing a novel CKM-based environment-aware ISAC framework. Different from the conventional environment-unaware ISAC approaches, which rely on the existence of a LoS path, the proposed method can exploit NLoS paths for sensing even when the LoS path is absent. The main contributions of this work are summarized as follows:

\begin{itemize}
\item \textbf{NLoS ISAC model}: We propose a novel system model for ISAC in NLoS scenarios without relying on those strong assumptions, such as the presence of a LoS path and reciprocal signal propagation where AoA equals AoD. Specifically, unlike existing models that usually assume signals return along the same path, failing to capture NLoS environments where signals may reflect off different scatterers, our model supports both reciprocal and non-reciprocal paths, allowing distinct AoA and AoD. This enables comprehensive path modeling, capturing all possible NLoS path combinations, enhancing applicability to diverse propagation scenarios and establishing a foundation for environment-aware NLoS ISAC.

\item \textbf{Fundamental principle of applying CKM to sensing}: This work elucidates the core principle of utilizing CKM, originally developed for communication \cite{zeng2021toward,zeng2024tutorial}, to enable environment-aware NLoS sensing, demonstrating that the same CKM can be used for both environment-aware NLoS sensing and communication. 
To this end, by treating sensing targets as virtual UE, CKM leverages location-specific channel knowledge (such as angles, delays) to infer sensing parameters. 
This is a non-trivial task, since a CKM built for communication learns one-way location-specific channel statistical priors, which cannot be directly applied to the two-way cascaded sensing channel.
Our primary contribution is the novel mathematical transformation framework that bridges this gap.
This approach utilizes environment priors embedded in CKM to exploit NLoS paths effectively, overcoming the limitations of environment-unaware methods that fail to map those estimated sensing parameters (such as multipath angles and delays) to target locations in NLoS scenarios.
\item \textbf{Sensing based on channel angle-delay map (CADM)}: We propose a particular type of CKM, named CADM, which tries to learn location-specific angle and delay distributions of channel multipath. For each path, CADM uses certain distributions like Gaussian distribution to characterize the angle and delay. The parameters of this distribution—mean and variance—are dependent on the specific location. As this dependency is usually highly complex and non-linear, making analytical modeling difficult, we employ a fully connected neural network (FCNN) to learn the mapping from location to these statistical parameters.
By treating the sensing target as a virtual UE, the communication channel knowledge distributions in the CADM, originally tied to UE location, are transformed into a sensing likelihood function corresponding to the sensing target's location.
Then, for target sensing, we perform maximum likelihood (ML) estimation. This process utilizes both the measured angle-delay data and the sensing likelihood function derived from CADM, and is optimized via gradient descent.
The proposed CADM-based method achieves environment-aware sensing by utilizing NLoS path information, where traditional LoS-based methods fail in pure NLoS environments.

\item \textbf{Extensive evaluation and superior performance}: Extensive simulations in the challenging NLoS environments validate the proposed method, demonstrating significant improvements in sensing accuracy compared to environment-unaware approaches, such as geometric triangulation. A rigorous performance analysis using the Cramér-Rao lower bound (CRLB) verifies the benefits of multipath exploitation in environment-aware sensing enabled by CKM.
\end{itemize}

The remainder of this paper is organized as follows. Section II presents the system model for the ISAC framework. Section III discusses the motivation for CKM-enabled sensing, while Section IV details the mechanisms for the proposed CADM-enabled NLoS sensing. Section V provides simulation results, and Section VI concludes the paper.

We adopt the following notations throughout this paper. Scalars are denoted by italic letters. Boldface lower- and upper-case letters denote vectors and matrices, respectively. $\mathbb{C}^{M \times N}$ and $\mathbb{R}^{M \times N}$ represent the spaces of $M \times N$ complex and real matrices, respectively. $ \|\mathbf{a}\| $ is the Euclidean norm of vector $ \mathbf{a} $. $\mathbf{A}^T$, $\mathbf{A}^*$, $\mathbf{A}^H$ denote the transpose, conjugate, and conjugate transpose of matrix $\mathbf{A}$, respectively. The operator $*$ represents convolution. The notation $\mathcal{N}(b; \mu, \sigma^2)$ represents a Gaussian distribution for variable $b$ with mean $\mu$ and variance $\sigma^2$. 
The notation $ \nabla_{\mathbf{x}} f(\mathbf{x})$ denotes the gradient of a scalar function $ f(\mathbf{x}) $ with respect to the vector $ \mathbf{x} $.
$ \text{blkdiag}(.) $ constructs a block diagonal matrix from its arguments.
The operator $ \lfloor  \cdot   \rfloor$ denotes the floor function, and the operator $ mod $ denotes the modulo operation.

\section{System Model}
\begin{figure}[h]
	\centering{\includegraphics[width=.48\textwidth]{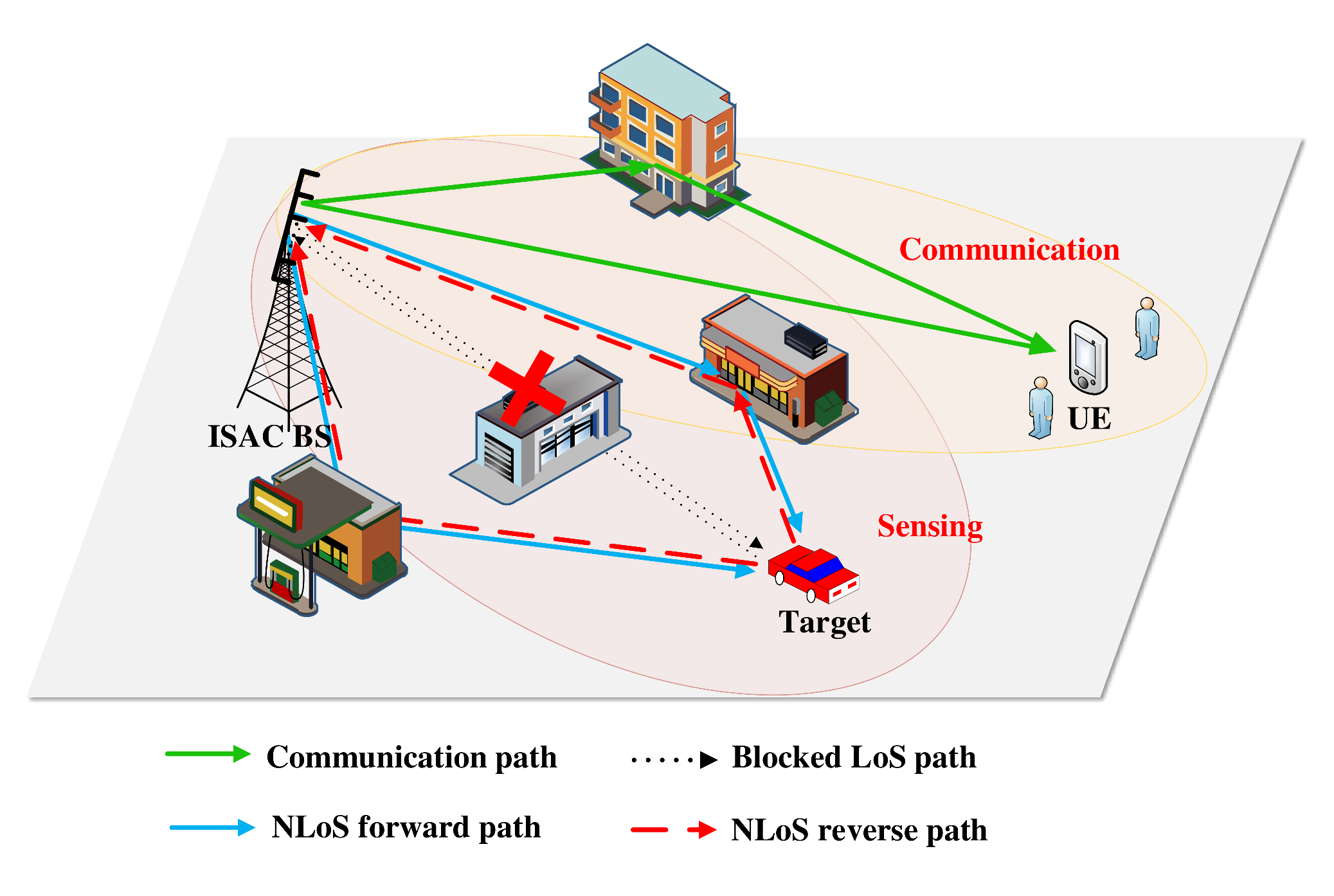}}  	
	\caption{An illustration of the ISAC system in NLoS environment, where the direct LoS link between the BS and the sensing target is blocked.} 
	\label{scene}  
\end{figure}

Consider an ISAC system, as shown in Fig. \ref{scene}, where a BS located at $\mathbf{b}=(x_{\mathbf{b}}, y_{\mathbf{b}})$ tries to communicate with a single-antenna UE and simultaneously provide sensing services.
The BS is equipped with $ N $ antennas for signal transmission and $N_{\text{Rx}}$ antennas for sensing signal reception. The UE is located at $\mathbf{q}=(x_{\mathbf{q}}, y_{\mathbf{q}})$.
While communicating with the UE, the ISAC system tries to estimate the location $\mathbf{x}=(x_{\mathbf{x}}, y_{\mathbf{x}})$ of the sensing object (e.g., vehicles or pedestrians) in the environment. We consider the mono-static sensing mode where the BS operates in full-duplex, utilizing the $N$ antennas for transmitting signals and the $N_{\text{Rx}}$ receive antennas to process the received echoes. Different from the conventional ISAC systems \cite{dai2025tutorial}, we focus on the challenging NLoS scenario where the LoS path between the BS and the target is blocked, relying solely on NLoS paths for target sensing, as illustrated in Fig. \ref{scene}.

\begin{figure*}[htbp]
	\centering{\includegraphics[width=.99\textwidth]{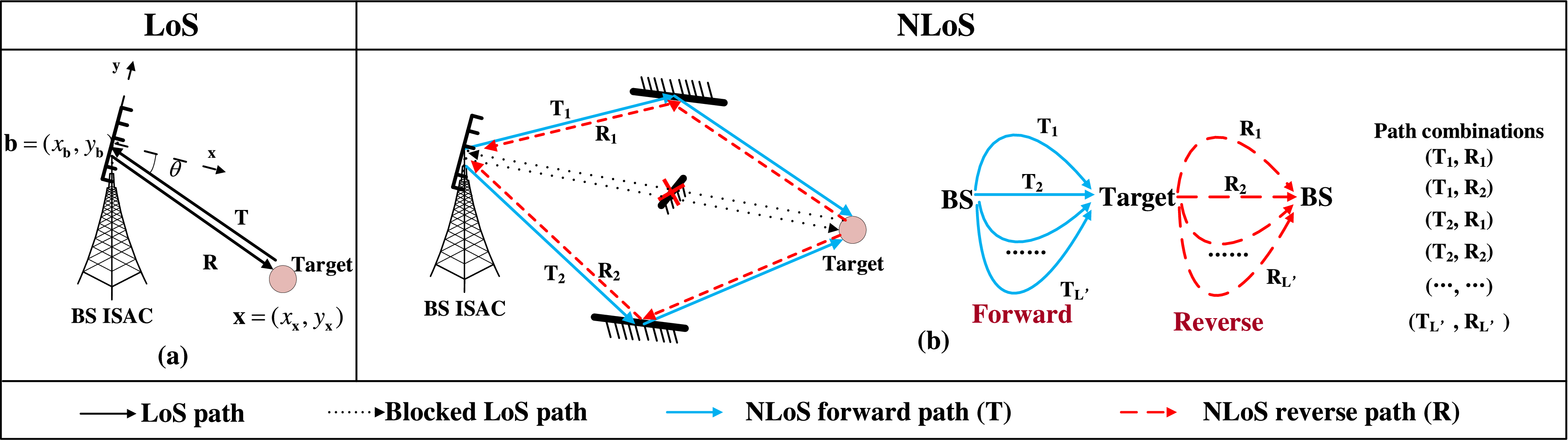}}  	
	\caption{An illustration of LoS versus NLoS mono-static sensing. } 
	\label{systemmodel}  
\end{figure*}

Traditional ISAC sensing models typically rely on the presence of a LoS path between the BS and the sensing target, assuming a single direct path with reciprocal signal propagation, i.e., the AoA equals the AoD \cite{dai2025tutorial}. In this case, as shown in Fig. \ref{systemmodel}(a), the signal travels along the same path for both the forward transmit and reverse reflected directions. In this case, the received sensing signal at the BS, denoted as $\mathbf{y}(t)$, is modeled as \cite{dai2025tutorial}
\begin{equation}
	\mathbf{y}(t) = \alpha \mathbf{a}_{r}(\theta) \mathbf{a}_{t}^H(\theta) \mathbf{s}(t - \tau) e^{j 2 \pi v t} +  \mathbf{n}(t),
	\label{losmo}
\end{equation}
where $\mathbf{s}(t)\in\mathbb{C}^{N\times1}$ is the transmitted signal, $\alpha$ is the complex gain of the path, $\mathbf{a}_{r}(\theta)\in \mathbb{C}^{N_\mathrm{Rx}\times1}$ and $\mathbf{a}_{t}(\theta)\in \mathbb{C}^{N\times1}$ are the steering vectors corresponding to the AoA $\theta$ for the $N_{\text{Rx}}$ receive antennas and the AoD $\theta$ for the $N$ transmit antennas, respectively. Besides, $\tau$ is the round-trip delay and $v$ is the Doppler shift, $\mathbf{n} \sim \mathcal{C N}(\mathbf{0}, \sigma^2 \mathbf{I}_{N_{\text{Rx}}})$ is the zero-mean circularly symmetric complex Gaussian (CSCG) noise vector with variance $\sigma^2$. 
Note that to gain the essential insights, we only focus on the echo signal from the sensing target by assuming that the clutter signal has been properly removed \cite{richards2005fundamentals}. 
In such a LoS scenario, it is further assumed for geometric consistency that the BS is equipped with a uniform linear array (ULA) oriented along the y-axis as shown in Fig. \ref{systemmodel}.
Therefore, the AoA/AoD $\theta$ and delay $\tau$ are directly related to the locations of the BS $ \mathbf{b} $ and the sensing target $ \mathbf{x} $, expressed as
\begin{equation}
	\left\{
	\begin{aligned}
		\theta &= \arctan\left(\frac{y_{\mathbf{b}} - y_{\mathbf{x}}}{x_{\mathbf{b}} - x_{\mathbf{x}}}\right) \\
		\tau &= \frac{2\|\mathbf{b} - \mathbf{x}\|}{c}
	\end{aligned}
	\right.,
	\label{conmoe}
\end{equation}
where $c$ is the speed of light. As a result, after estimating the parameters $ \hat{\theta} $ and $ \hat{\tau} $ based on the echo signal $ \mathbf{y}(t)  $ in (\ref{losmo}) with various algorithms reviewed in \cite{dai2025tutorial}, the location of the sensing target $ \mathbf{x} $ can be directly obtained as
	\begin{equation}
		\left\{
		\begin{aligned}
			x_{\mathbf{x}} &= \frac{\hat{\tau} c}{2}\cos(\hat{\theta})+x_{\mathbf{b}} \\
			y_{\mathbf{x}} &= \frac{\hat{\tau} c}{2}\sin(\hat{\theta})+y_{\mathbf{b}}
			\label{xyco}
		\end{aligned}
		\right..
	\end{equation}

However, the above result is only applicable when the LoS link between the BS and sensing targets exists. In the challenging NLoS scenario where the LoS link is blocked, as illustrated in Fig. \ref{systemmodel}(b), the signal transmitted by the BS arrives at the sensing target via forward paths with reflections, diffractions, or scattering, is reflected by the sensing target, and then returns to the BS via reverse paths. In this case, there exist multiple paths between the BS and the target, each characterized by distinct AoA, AoD, delays, and Doppler shifts. The received echo signal at the BS, denoted as $\mathbf{y}(t)$, is thus expressed as
\begin{equation}
	\begin{aligned}
		\mathbf{y}(t)=&\sum_{l=1}^{L}\alpha_l \mathbf{a}_{r}(\theta_l)\mathbf{a}_{t}^H(\phi_l)\mathbf{s}(t-\tau_l)e^{j2\pi v_l t}+\mathbf{n}(t),
	\end{aligned}
	\label{eqmonosignal}
\end{equation}
where $L$ denotes the number of combinations of the round-trip paths, $\alpha_l$ is the complex gain of the $l$-th path, $\mathbf{a}_{r}(\theta_l)$ and $\mathbf{a}_{t}(\phi_l)$ are the steering vectors for the AoA $\theta_l$ and AoD $\phi_l$, respectively, $\tau_l$ is the delay, and $v_l$ is the Doppler shift.

Note that in NLoS scenarios specified in (\ref{eqmonosignal}), the signal may travel along paths formed by different combinations of the forward transmit and reverse reception directions, denoted as $ T $ and $ R $, as shown in Fig. \ref{systemmodel}(b). Note that different from the LoS sensing model in (\ref{losmo}), where the signal returns along the same path it took outbound \cite{dai2025tutorial}, the NLoS model in (\ref{eqmonosignal}) allows $\theta_l$ and $\phi_l$ to be different. For example, the signal can depart along path $ T_1 $ and return via path $ R_2 $, or depart along $ T_2 $ and return via $ R_1 $, leading to distinct AoA and AoD ($\theta_l \neq \phi_l$). Therefore, the model (\ref{eqmonosignal}) captures all possible echo signals that can return to the BS. For instance, if there are $L'$ forward transmit paths and $L'$ reverse reflected paths, and we only consider $L'$ reciprocal paths (e.g., $ T_1 $ to $ R_1 $, $ T_2 $ to $ R_2 $, ..., $ T_{L'} $ to $ R_{L'} $), the received signal is written as
\begin{equation}
	\begin{aligned}
		\mathbf{y}(t)=\sum_{l'=1}^{L'}\alpha_{l'} \mathbf{a}_{r}(\theta_{l'})\mathbf{a}_{t}^H(\theta_{l'})\mathbf{s}(t-\tau_{l'})e^{j2\pi v_{l'} t}+\mathbf{n}(t),
	\end{aligned}
	\label{eqmonosignaltra}
\end{equation}
which ignores a significant number of potential paths.
In contrast, the proposed model in (\ref{eqmonosignal}) accounts for all $L=L'^2$ possible path combinations, and the received signal in (\ref{eqmonosignal}) can be written as 

\begin{equation}
	\begin{aligned}
				\mathbf{y}(t)=&\sum_{l_T=1}^{L'}\sum_{l_R=1}^{L'}\alpha_{l_T,l_R} \mathbf{a}_{r}(\theta_{l_R})\mathbf{a}_{t}^H(\phi_{l_T})\mathbf{s}(t-\tau_{l_T, l_R})\\&\times e^{j2\pi v_{l_T, l_R} t}+\mathbf{n}(t),
	\end{aligned}
	\label{eqm}
\end{equation}
where $ \alpha_{l_T,l_R}, \tau_{l_T,l_R} $ and $ v_{l_T,l_R} $ respectively denote the complex gain, delay, and Doppler shift for the path combination $ (l_T,l_R) $. This formula includes all the $L=L'^2$ combinations of forward and reverse paths.

To perform target sensing in NLoS scenarios, it is essential to extract target-related channel knowledge from the received signal $\mathbf{y}(t)$. Specifically, target-related channel parameters, such as AoD $\phi_{l_T}$, AoA $\theta_{l_R}$, delay $\tau_{l_T, l_R}$, and Doppler shift $v_{l_T, l_R}$ for each path combination $ (l_T,l_R) $, can be estimated based on the echo signal $ \mathbf{y}(t) $ using methods like multiple signal classification (MUSIC) or estimation of signal parameters via rotational invariance techniques (ESPRIT) \cite{dai2025tutorial,schmidt1986multiple,roy1989esprit}. For each of the $L=L'^2$ paths in the NLoS scenario, the channel knowledge vector for the path combination $ (l_T,l_R) $, denoted as $\mathbf{z}_{{l_T},l_R} = [\phi_{l_T}, \theta_{l_R}, \tau_{l_T,l_R}]^T$, captures the AoD $\phi_{l_T}$, AoA $\theta_{l_R}$, and delay $ \tau_{l_T,l_R}$. While the Doppler shift $v_{l_T,l_R}$ is estimated for potential use in dynamic scenarios, our focus on target location estimation leads us to aggregate only $\phi_{l_T}$, $\theta_{l_R}$, and $\tau_{l_T,l_R}$ across all paths. The observed channel knowledge for all $L$ paths is thus represented as $\mathbf{z} \in \mathbb{R}^{3L \times 1}$, formed by concatenating the vectors as
\begin{equation}
	\mathbf{z} = [\mathbf{z}_{1,1}^T, \mathbf{z}_{1,2}^T, \ldots, \mathbf{z}_{{l_T},l_R} ^T,...,\mathbf{z}_{{L'},L'} ^T]^T\in\mathbb{R}^{3L\times1},
\end{equation}
where each $\mathbf{z}_{{l_T},l_R}$ contributes three parameters, resulting in a total of $3L$ elements in $\mathbf{z}$.

To ultimately achieve the sensing goal, these parameters in $\mathbf{z}$ must be mapped to the location $\mathbf{x}$ of the sensing target. In LoS scenarios, this mapping can be directly achieved with the closed-form expression (\ref{xyco}).
However, such a mapping is no longer valid for the challenging scenario where LoS is absent, since NLoS paths depend on the environment, denoted as $E$, as shown in Fig. \ref{scene}. Therefore, when only NLoS paths are available, without any prior knowledge about the environment, determining the sensing target location $\mathbf{x}$ based on the sensed channel parameters $\mathbf{z}$ is an underdetermined problem. In fact, even in mixed scenarios with both LoS and NLoS paths, conventional methods typically employ techniques such as NLoS mitigation or path selection to isolate the LoS path, discarding NLoS paths that contain valuable target-related information \cite{marano2010nlos}. This results in information loss, limiting the sensing accuracy. To utilize the target-related information in NLoS paths, which depends on $E$, an environment-aware sensing model that incorporates $E$ is necessary to leverage both LoS (if it exists) and NLoS paths for sensing target localization.

To achieve the above goal, this paper first develops an environment-aware observation model as
\begin{equation}
	\hat{\mathbf{z}} = f(\mathbf{x}, E) + \mathbf{e},
	\label{ob-model}
\end{equation}
where $f(\cdot)$ is an abstract representation of the mapping from the sensing target location $\mathbf{x}$ and environment $ E $ to the observed channel knowledge, denoted as $\hat{\mathbf{z}}$.
In the LoS case, where $E = \emptyset$, the mapping $f(\cdot)$ degenerates to the closed-form expression in (\ref{conmoe}).
The vector $\mathbf{e} \in \mathbb{R}^{3L \times 1}$ models the errors associated with the angle and delay estimations, assumed to follow an independent multivariate Gaussian distribution $\mathbf{e} \sim \mathcal{N}(\mathbf{0}, \mathbf{\Sigma}_e)$, with
\begin{equation}
	\mathbf{\Sigma}_e = \text{blkdiag}(\sigma_{\phi}^2,\sigma_{\theta}^2,\sigma_{\tau}^2),
\end{equation}
where $ \sigma_{\phi}^2 $, $\sigma_{\theta}^2$ and $\sigma_{\tau}^2$ are the variances of the AoD, AoA and delay estimation errors, respectively.

Given the observation model in (\ref{ob-model}), the sensing target location $\mathbf{x}$ is to be estimated. In NLoS scenarios, the environment-dependent nature of $f(\cdot)$ makes it challenging to directly infer $ \mathbf{x} $ based on $ \hat{\mathbf{z}} $. To address this, we adopt a general probabilistic framework using ML estimation, a promising method that handles both the complexity of NLoS scenarios and the uncertainty introduced by estimation errors. The ML estimate of the target location $ \mathbf{x} $ based on $ \hat{\mathbf{z}} $ can be expressed as
\begin{equation}
	\hat{\mathbf{x}} = \mathop{\arg\max}_{\mathbf{x}}\ P(\hat{\mathbf{z}}|\mathbf{x}) = \mathop{\arg\min}_{\mathbf{x}}\ -\ln P(\hat{\mathbf{z}}|\mathbf{x}),
	\label{maxP}
\end{equation}
where $P(\hat{\mathbf{z}} | \mathbf{x})$ represents the likelihood function of the observed channel knowledge $ \hat{\mathbf{z}} $. With Gaussian error $\mathbf{e}$, the log-likelihood function can be expressed as 
\begin{equation}
	\ln P(\hat{\mathbf{z}}|\mathbf{x}) =\ln\mathcal{N}(\hat{\mathbf{z}};f(\mathbf{x}, E),\mathbf{\Sigma}_e).
		\label{log-likelihood}
\end{equation}

The problem (\ref{maxP}) can be solved by classic optimization methods such as the gradient descent method. The update rule for the target location at the $(k+1)$th iteration is given by
\begin{equation}
	\mathbf{x}^{(k+1)} = \mathbf{x}^{(k)} - \eta \left. \nabla_{\mathbf{x}} \left(-\ln P(\hat{\mathbf{z}} | \mathbf{x})\right) \right|_{\mathbf{x}^{(k)}},
	\label{xk}
\end{equation}
where $\eta$ denotes the step size, and $\nabla_{\mathbf{x}} (-\ln P(\hat{\mathbf{z}} | \mathbf{x}))$ represents the gradient with respect to $\mathbf{x}$. Based on (\ref{log-likelihood}), this gradient can be expressed as
\begin{equation}
	\nabla_{\mathbf{x}} \left(-\ln P(\hat{\mathbf{z}} | \mathbf{x})\right) = -\mathbf{J}^T \mathbf{\Sigma}_e^{-1} \left(\hat{\mathbf{z}} - f(\mathbf{x}, E)\right),
\end{equation}
where $\mathbf{J} = \frac{\partial f}{\partial \mathbf{x}} \in \mathbb{R}^{{3L} \times 2}$ denotes the Jacobian matrix of the mapping function $f(\cdot)$ with respect to $\mathbf{x}$, given by
\begin{equation}
	\mathbf{J} = \frac{\partial f}{\partial \mathbf{x}} = \begin{bmatrix}
		\frac{\partial f_1}{\partial x_{\mathbf{x}}} & \frac{\partial f_1}{\partial y_{\mathbf{x}}} \\
		\frac{\partial f_2}{\partial x_{\mathbf{x}}} & \frac{\partial f_2}{\partial y_{\mathbf{x}}} \\
		\vdots & \vdots \\
		\frac{\partial f_{3L}}{\partial x_{\mathbf{x}}} & \frac{\partial f_{3L}}{\partial y_{\mathbf{x}}}
	\end{bmatrix}.
	\label{jacob}
\end{equation}

From (\ref{jacob}), the critical role of the mapping function $f(\cdot)$ is evident. To estimate $\mathbf{x}$ using the gradient-based approach, $f(\cdot)$ must be explicitly known and differentiable with respect to $\mathbf{x}$. This requirement highlights the importance of modeling the relationship between the observed channel knowledge $ \hat{\mathbf{z}} $ and the target location $ \mathbf{x} $ in practical NLoS sensing systems.

To provide a clearer insight into the benefits of environment-aware sensing, the sensing performance is quantified using CRLB, which provides a theoretical lower bound on the variance of unbiased parameter estimation \cite{kay1993fundamentals}. The CRLB is derived based on the Fisher information matrix, which quantifies the amount of information that the observed data $\hat{\mathbf{z}}$ carries about the parameter $\mathbf{x}$. For the observation model in (\ref{ob-model}) with the log-likelihood function given by (\ref{log-likelihood}), the Fisher information matrix (FIM) $\mathbf{I}(\mathbf{x})$ can be written as
\begin{equation}
	\begin{aligned}
		\mathbf{I}(\mathbf{x}) &= \mathbb{E}\left[\left(\nabla_{\mathbf{x}} \ln P(\hat{\mathbf{z}} \mid \mathbf{x})\right)\left(\nabla_{\mathbf{x}} \ln P(\hat{\mathbf{z}} \mid \mathbf{x})\right)^T\right] \\
		&= \mathbb{E}\left[\mathbf{J}^T \mathbf{\Sigma}_e^{-1} (\hat{\mathbf{z}} - f(\mathbf{x}, E)) (\hat{\mathbf{z}} - f(\mathbf{x}, E))^T \mathbf{\Sigma}_e^{-1} \mathbf{J}\right] \\
		&= \mathbf{J}^T \mathbf{\Sigma}_e^{-1} \mathbb{E}\left[(\hat{\mathbf{z}} - f(\mathbf{x}, E)) (\hat{\mathbf{z}} - f(\mathbf{x}, E))^T\right] \mathbf{\Sigma}_e^{-1} \mathbf{J}.
	\end{aligned}
\label{fishermatrix}
\end{equation}
Since $\hat{\mathbf{z}} - f(\mathbf{x}, E) = \mathbf{e}$ and $\mathbf{e} \sim \mathcal{N}(\mathbf{0}, \mathbf{\Sigma}_e)$, the expectation simplifies to $\mathbb{E}\left[(\hat{\mathbf{z}} - f(\mathbf{x}, E)) (\hat{\mathbf{z}} - f(\mathbf{x}, E))^T\right] = \mathbb{E}\left[\mathbf{e} \mathbf{e}^T\right] = \mathbf{\Sigma}_e$, so the Fisher information matrix becomes
\begin{equation}
	\mathbf{I}(\mathbf{x}) = \mathbf{J}^T \mathbf{\Sigma}_e^{-1} \mathbf{\Sigma}_e \mathbf{\Sigma}_e^{-1} \mathbf{J} = \mathbf{J}^T \mathbf{\Sigma}_e^{-1} \mathbf{J}.
\end{equation}
The CRLB for the estimation of $\mathbf{x}$ is then the inverse of the Fisher information matrix, given by
\begin{equation}
	\text{CRLB} = \mathbf{I}(\mathbf{x})^{-1} = \left(\mathbf{J}^T \mathbf{\Sigma}_e^{-1} \mathbf{J}\right)^{-1}.
	\label{varx}
\end{equation}
Since $\mathbf{J}$ is of dimension ${3L} \times 2$, and assuming independent observations with a constant $\mathbf{\Sigma}_e$, the Fisher information matrix $\mathbf{J}^T \mathbf{\Sigma}_e^{-1} \mathbf{J}$ scales with $3L$ \cite{kay1993fundamentals}. Therefore, the CRLB scales inversely with $3L$, which can be written as
\begin{equation}
	\text{CRLB} \sim O\left(\frac{1}{3L}\right).
	\label{Evarx}
\end{equation}

Environment-aware sensing, as derived from (\ref{Evarx}), reveals that sensing performance scales proportionally with both the number of paths $ L $ and the amount of channel knowledge available per path. Firstly, the modeling in (\ref{eqmonosignal}) increases the number of usable paths from $L'$ to $L'^2$ by capturing all possible path combinations, thereby enriching the multipath information. Additionally, different from the LoS modelling with the assumption $\theta = \phi$, each path provides three channel knowledge parameters (AoD, AoA, and delay), compared to only two in prior models (\ref{conmoe}), further enhancing the data available for sensing. Secondly, the observation underscores a key insight for environment-aware sensing---while conventional environment-unaware sensing methods typically suffer from degraded performance in complex environments due to increased interference, environment-aware sensing counterintuitively benefits from such complexity. Specifically, if those NLoS paths can be exploited, a more complex environment provides more NLoS multipath information, which means increased $ L $, thereby enhancing sensing performance.

However, the practical implementation of environment-aware sensing based on (\ref{xk}) is highly non-trivial. The first major difficulty lies in acquiring an accurate representation of the environment $E$ and formulating its mathematical model. The second challenge is that even if $E$ is known, the mapping function $f(\cdot)$ remains difficult to determine precisely. Potential approaches, such as solving Maxwell's equations or employing ray tracing techniques, suffer from prohibitively high computational complexity, rendering them unsuitable for real-time sensing applications. Furthermore, these approaches cannot guarantee differentiability, which is essential for gradient-based optimization methods in ML estimation as in (\ref{xk}).

Fortunately, the recently proposed concept of CKM offers a promising solution to enable efficient environment-aware sensing \cite{zeng2021toward,zeng2024tutorial}. Initially developed to capture prior channel knowledge between BS and UE locations for enhancing communication performance, CKM has been recognized as a versatile tool that can also significantly benefit sensing applications. By leveraging the spatial and environment insights embedded in CKM, it is feasible to address the challenges for modeling the environment $E$ and the mapping function $f(\cdot)$. In the following sections, the motivation and mechanisms for using the same CKM designed for communication purposes to enable NLoS sensing will be elaborated.

\section{Environment-aware ISAC via CKM}
CKM was originally developed to infer prior channel knowledge based on UE locations for communication purposes. In this context, the sensing target location $\mathbf{x}$ was not involved. CKM offers a method to bypass the requirement for explicit modeling of $E$ or $f(\cdot)$, circumventing the complexities associated with environmental characterization and function derivation. Extending this framework to map sensing target locations $\mathbf{x}$ to channel knowledge $\mathbf{z}$ faces new challenges and opportunities, which motivates the exploration of CKM-enabled sensing in this work. In the following, we present the motivation behind this extension, detailing its core mechanism for enhancing environment-aware sensing capabilities.

Let $\mathbf{Q} = [\mathbf{q}_1, \ldots, \mathbf{q}_Q] \in \mathbb{R}^{2 \times Q}$ and $\mathbf{W} = [\mathbf{w}_1, \ldots, \mathbf{w}_Q] \in \mathbb{R}^{D \times Q}$ denote the $Q$ historically visited UE locations and their corresponding observed $D$-dimensional channel knowledge data, respectively \cite{zeng2024tutorial}. These location-channel knowledge pairs $(\mathbf{Q}, \mathbf{W})$ are utilized to learn prior channel knowledge, forming the basis of CKM. CKM can thus be expressed as
\begin{equation}
	(\mathbf{Q},\mathbf{W})\Longrightarrow \mathcal{M}: \mathbb{R}^{2\times1}\rightarrow \mathbb{R}^{D\times1},
	\label{ckmmap}
\end{equation}
where $\mathcal{M}$ represents the CKM, mapping a two-dimensional UE location vector to a channel vector of dimension $D$. 
CKM exploits these historical pairs $ (\mathbf{Q},\mathbf{W}) $ to infer new channel knowledge without requiring an explicit characterization of either the mapping function $f(\cdot)$ or the environment $E$.
Without loss of generality, CKM captures the statistical relationship between UE locations $ \mathbf{q} $ and channel knowledge $ \mathbf{w} $, represented as a conditional probability distribution $P_{\mathcal{M}}(\mathbf{w}|\mathbf{q})$. This relationship is learned from the historical data $(\mathbf{Q}, \mathbf{W})$ through various methods, such as interpolation, machine learning, or other data-driven techniques, depending on the specific implementation of CKM. 

In communication applications, CKM utilizes prior information about UE location $P(\mathbf{q})$ to infer prior channel knowledge $	\hat{\mathbf{w}}$ based on the learned $P_{\mathcal{M}}(\mathbf{w}|\mathbf{q})$, which can be written as
\begin{equation}
\hat{\mathbf{w}} = \mathop{\arg\max}_{\mathbf{w}} \int P_{\mathcal{M}}(\mathbf{w}|\mathbf{q}) P(\mathbf{q}) d\mathbf{q}.
\end{equation}
The inferred channel knowledge $\hat{\mathbf{w}}$ is subsequently used to optimize various aspects of communication systems, such as beamforming, resource allocation, and interference management \cite{zeng2021toward,zeng2024tutorial,jiang2025interference, wang2024channel,wu2023environment}. However, leveraging CKM to enhance sensing capabilities, particularly to map sensing target locations $\mathbf{x}$ to channel knowledge $	\hat{\mathbf{w}}$, remains unexplored.

In principle, if the channel knowledge type is appropriately designed and a sufficient number of locations have been sampled, the historical data $(\mathbf{Q}, \mathbf{W})$ holds the potential to reconstruct the environment $E$, which encapsulates the physical characteristics affecting signal propagation. Such a reconstruction can be abstractly written as \cite{zeng2024tutorial}
\begin{equation}
	E=g(\mathbf{Q},\mathbf{W}),
\end{equation}
where $g(\cdot)$ denotes a function that reconstructs $E$ based on the historical data. Substituting this into the sensing observation model (\ref{ob-model}), the observed channel knowledge can be reformulated as
\begin{equation}
	\begin{aligned}
				\hat{\mathbf{z}}&=f(\mathbf{x},g(\mathbf{Q},\mathbf{W}))+\mathbf{e}\\
			&=h(\mathbf{x};\mathbf{Q},\mathbf{W})+\mathbf{e},
			\label{hfun}
	\end{aligned}
\end{equation}
where $h(\cdot)$ is the function taking the composition of functions $ f(\cdot)  $ and $ g(\cdot) $. From (\ref{hfun}), the historical data $(\mathbf{Q}, \mathbf{W})$ can effectively replace the requirement to explicitly reconstruct the environment $E$.

CKM is obtained by learning from the historical data $(\mathbf{Q}, \mathbf{W})$, represented by the mapping $\mathcal{M}$ as shown in (\ref{ckmmap}). 
Different from explicitly reconstructing the physical environment $ E $ from the historical data $(\mathbf{Q}, \mathbf{W})$, which is usually computationally prohibitive and unnecessary, CKM serves as an implicit, data-driven representation of the environment $E$.
It directly learns the functional ``location-to-channel" mapping from historical data, effectively capturing the necessary environment prior knowledge without the overhead of an explicit physical map reconstruction. Therefore, the observation model (\ref{hfun}) can be further written as
\begin{equation}
			\hat{\mathbf{z}}=h(\mathbf{x};\mathcal{M})+\mathbf{e}.
\end{equation}
With the CKM $\mathcal{M}$ constructed, this formulation suggests the potential to infer the relationship between sensing target location $\mathbf{x}$ and channel knowledge observation $\hat{\mathbf{z}}$, achieving environment-aware sensing in complex environments. Since a key step of environment-aware sensing is to obtain the likelihood $P(\hat{\mathbf{z}}|\mathbf{x})$ in (\ref{xk}), CKM-enabled sensing is to infer the \emph{target location-channel knowledge} relationship $P(\hat{\mathbf{z}}|\mathbf{x})$ based on the \emph{UE location-channel knowledge} relationship $P_{\mathcal{M}}(\mathbf{w}|\mathbf{q})$, i.e.,
\begin{equation}
	P_{\mathcal{M}}(\mathbf{w}|\mathbf{q}) \rightarrow P(\hat{\mathbf{z}}|\mathbf{x}).
	\label{transition}
\end{equation}

\begin{figure}[htbp]
	\centering{\includegraphics[width=.48\textwidth]{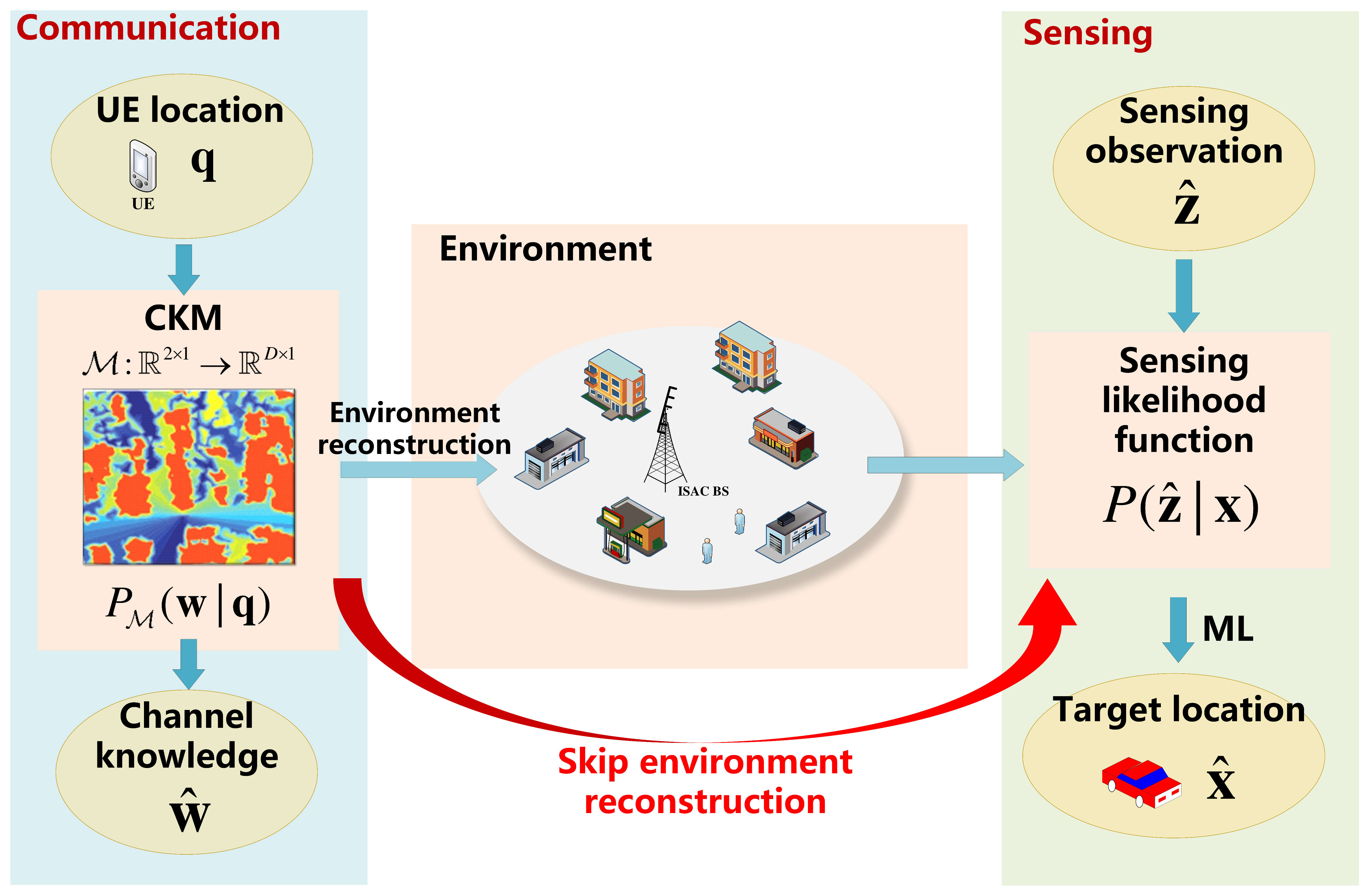}}  	
	\caption{An illustration of CKM-enabled ISAC that utilizes the same CKM $ \mathcal{M} $ developed for communication to achieve environment-aware sensing as well. The CKM, by learning the ``location-to-channel" mapping ($P_{\mathcal{M}}(\mathbf{w}|\mathbf{q})$) from data, serves as an implicit representation of the environment $ E $. This allows the framework to ``skip" the computationally expensive step of explicit physical environment reconstruction and directly provide the sensing likelihood $P(\hat{\mathbf{z}}|\mathbf{x})$ for sensing target localization.} 
	\label{motivation}  
\end{figure}

The transition in (\ref{transition}) represents a pivotal advancement, transforming the CKM—originally developed to infer the channel knowledge $ \mathbf{w} $ with UE location $ \mathbf{q} $—into a tool for inferring sensing channel observation $ \hat{\mathbf{z}} $ for sensing target location $ \mathbf{x} $ by using one single CKM, as shown in Fig. \ref{motivation}. 
This embodies the core theme of this paper, i.e., \emph{you may use the same CKM for both environment-aware NLoS sensing and communication}, eliminating the need for additional sensing-specific CKM construction.
The detailed methodologies for achieving this transformation will be investigated in the following sections.

\section{CKM-Enabled Environment-Aware NLoS Sensing}

\begin{figure*}[htbp]
	\centering{\includegraphics[width=.96\textwidth]{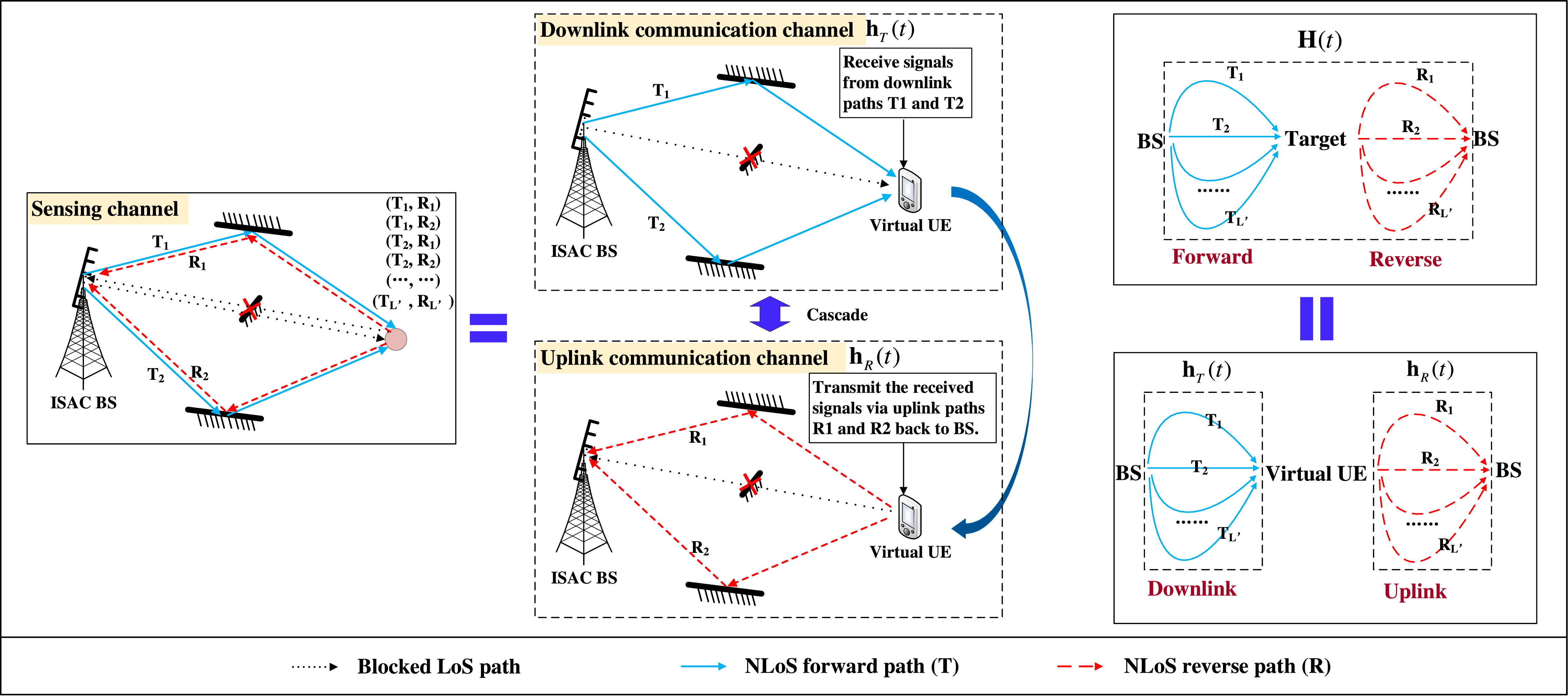}}  	
	\caption{An illustration of the relationship between sensing response and communication channel.} 
	\label{method}  
\end{figure*}

To explain how the CKM designed for communication purposes can be reused for sensing, it is essential to first reveal the relationship between the sensing response and the communication channel, followed by outlining the specific methods for implementation.

\subsection{Relationship between sensing response and communication channel}
A target that reflects signals in all directions is analogous to a virtual communication UE at the same location. This analogy arises from the physical similarity in signal interaction: the target receives the transmitted signal from the BS, attenuates it with a reflection coefficient $\beta$, and reflects it back to the BS. This process mirrors the behavior of a virtual UE that receives the downlink signal and retransmits it via an uplink channel, as illustrated in Fig. \ref{method}.

Let $ \mathbf{h}_T(t) $ and $ \mathbf{h}_R(t) $ denote the downlink and uplink channels between the BS and the virtual UE (target) respectively, given by
\begin{equation}
	\begin{aligned}
		\mathbf{h}_T(t)&=\sum_{l_T=1}^{L'}\alpha_{l_T}\mathbf{a}(\theta_{l_T})\delta(t-\tau_{l_T})\\
		\mathbf{h}_R(t)&=\sum_{l_R=1}^{L'}\alpha_{l_R}\mathbf{a}(\theta_{l_R})\delta(t-\tau_{l_R})
	\end{aligned}
	\label{commchannel}
\end{equation}
with $\alpha_{l_T}$ and $\alpha_{l_R}$ representing the complex gain of the $l_T$th forward and $l_R$th reverse paths, respectively, $\tau_{l_T}$ and $\tau_{l_R}$ indicating the delays, and $L'$ signifying the number of paths for each channel.
To model the complete sensing channel, a single path combination is first considered. The downlink signal received at the virtual UE (target) from the $l_T$-th path is given by
\begin{equation}
	y_{T,l_T}(t)=\alpha_{l_T}\mathbf{a}^H(\theta_{l_T})\mathbf{s}(t-\tau_{l_T}),
\end{equation}
where $\mathbf{s}(t) \in \mathbb{C}^{N \times 1}$ is the transmitted signal from the BS. This signal is then reflected by the target and propagates back to the BS via a reverse path, which is denoted as the $l_R$th path. This physical interaction, which couples the incoming path $l_T$ to the outgoing path $l_R$, is characterized by an angle-dependent reflection coefficient $\beta(\theta_{l_T}, \theta_{l_R})$. This coefficient represents the target's radar cross-section (RCS) dependent on the incident angle from path $l_T$ and reflected angle to path $l_R$.

The signal component that is reflected from path $l_T$ onto path $l_R$ can thus be expressed as $\beta(\theta_{l_T}, \theta_{l_R}) y_{T,l_T}(t)$. This reflected signal component then propagates through the $l_R$-th uplink path, whose impulse response is $\mathbf{h}_{R,l_R}(t) = \alpha_{l_R}\mathbf{a}(\theta_{l_R})\delta(t-\tau_{l_R})$.
The received signal component $\mathbf{y}_{l_R, l_T}(t)$ at the BS for this single path combination is the convolution of the uplink path response and the reflected signal component, which can be written as
\begin{equation}
	\begin{aligned}
		\mathbf{y}_{l_R, l_T}(t) &= \mathbf{h}_{R,l_R}(t) \ast \big( \beta(\theta_{l_T}, \theta_{l_R}) y_{T,l_T}(t) \big) \\
		&= \big( \alpha_{l_R}\mathbf{a}(\theta_{l_R})\delta(t-\tau_{l_R}) \big) \\
		& \quad \ast \big( \beta(\theta_{l_T}, \theta_{l_R}) \alpha_{l_T}\mathbf{a}^H(\theta_{l_T})\mathbf{s}(t-\tau_{l_T}) \big) \\
		&= \beta(\theta_{l_T}, \theta_{l_R}) \alpha_{l_R} \alpha_{l_T} \mathbf{a}(\theta_{l_R}) \\
		& \quad \times \mathbf{a}^H(\theta_{l_T}) \mathbf{s}\big(t - (\tau_{l_R} + \tau_{l_T})\big).
	\end{aligned}
\end{equation}
The total signal $\mathbf{y}(t)$ received at the BS is the superposition of all $L'^2$ possible path combinations, which can be written as
\begin{equation}
	\begin{aligned}
		\mathbf{y}(t) &= \sum_{l_R=1}^{L'}\sum_{l_T=1}^{L'} \mathbf{y}_{l_R, l_T}(t) + \mathbf{n}(t) \\
		&= \sum_{l_R=1}^{L'}\sum_{l_T=1}^{L'} \beta(\theta_{l_T}, \theta_{l_R}) \alpha_{l_R} \alpha_{l_T} \mathbf{a}(\theta_{l_R}) \\
		& \quad \times \mathbf{a}^H(\theta_{l_T}) \mathbf{s}\big(t - (\tau_{l_R} + \tau_{l_T})\big) + \mathbf{n}(t) \\
		&= \mathbf{H}(t)\ast\mathbf{s}(t)+\mathbf{n}(t),
	\end{aligned}
	\label{y3}
\end{equation}
where $\mathbf{H}(t)$ is the total sensing response matrix, characterizing the signal propagation from transmission to reflection and back to reception. From (\ref{y3}), $\mathbf{H}(t)$ is defined as
\begin{equation}
	\begin{aligned}
		\mathbf{H}(t) &= \sum_{l_R=1}^{L'} \sum_{l_T=1}^{L'} \beta(\theta_{l_T}, \theta_{l_R}) \alpha_{l_R} \alpha_{l_T} \mathbf{a}(\theta_{l_R}) \\
		& \quad \times \mathbf{a}^H(\theta_{l_T}) \delta\big(t - (\tau_{l_R} + \tau_{l_T})\big).
	\end{aligned}
	\label{hcas}
\end{equation}
This derivation shows that the sensing response $\mathbf{H}(t)$ is a superposition of all $L'^2$ path combinations, where each path's contribution is scaled by its unique angle-dependent RCS, $\beta(\theta_{l_T}, \theta_{l_R})$. This demonstrates that the sensing response matrix can be inferred from the communication channel, by treating the sensing target as the virtual UE at the same location. Therefore, the prior knowledge of the communication channel between the BS and the virtual UE (i.e., the target) serves as a valuable source of prior information, facilitating sensing tasks by leveraging the established channel characteristics.

By comparing the cascaded channel model (\ref{hcas}) with the general sensing model in (\ref{eqmonosignal}) and (\ref{eqm}), the explicit relationship between their respective parameters can be revealed. The comparison is performed in two logical steps.
First, the derived sensing response $ \mathbf{H}(t) $ in (\ref{hcas}) is compared with the detailed NLoS sensing model presented in (\ref{eqm}). Since both expressions are formulated as a double summation over the forward paths ($ l_T $) and reverse paths ($ l_R $), we can perform a term-by-term comparison for each $ (l_T, l_R) $ pair. This yields the following equivalences for the path parameters:
\begin{itemize}
	\item \textbf{Sensing AoD}: $ \phi_{l_T} $ from (\ref{eqm}) corresponds to the communication path AoD $ \theta_{l_T} $ from (\ref{hcas}).
	\item \textbf{Sensing AoA}: $ \theta_{l_R} $ from (\ref{eqm}) corresponds to the communication path AoA $ \theta_{l_R} $ from (\ref{hcas}).
	\item \textbf{Sensing Delay}: The total delay $ \tau_{l_T,l_R} $ for the combined path in (\ref{eqm}) corresponds to the sum of the communication path delays $ \tau_{l_T}+\tau_{l_R} $ from (\ref{hcas}).
\end{itemize}
Second, recall that the model in (\ref{eqm}) is an expanded representation of the general sensing model in (\ref{eqmonosignal}). The connection between the single path index $ l $ in (\ref{eqmonosignal}) and the index pair $ (l_T, l_R) $ in (\ref{eqm}) is given by the mapping $l=L'(l_T-1)+l_R$.
Therefore, by combining these steps, the definitive relationship between the parameters of the $ l $th sensing path and its constituent communication paths can be written as
\begin{equation}
	\begin{aligned} 
		\phi_l &= \phi_{l_T} = \theta_{l_T},\\
		\theta_l &= \theta_{l_R} = \theta_{l_R},\\
		\tau_l &= \tau_{l_T,l_R} = \tau_{l_T}+\tau_{l_R},
	\end{aligned}
\label{equa}
\end{equation}
where, for each row, the first, second, and third terms correspond to the parameters from (\ref{eqmonosignal}), (\ref{eqm}), and (\ref{hcas}), respectively.
Note that this extrapolation is universally applicable since the underlying relationships in (\ref{equa}) are independent of the target's physical properties. As shown in (\ref{y3}), the angle-dependent reflection coefficient $\beta(\theta_{l_T}, \theta_{l_R})$ models the target's specific scattering characteristics (e.g., shape or RCS) and acts as a component of the total complex gain $\alpha_{l_T, l_R}$. However, it does not change the fundamental geometric parameters (angles $\theta_{l_T}, \theta_{l_R}$ and delays $\tau_{l_T}, \tau_{l_R}$), which are dictated solely by the propagation environment. Therefore, these relationships in (\ref{equa}) hold true for any passive sensing target at that location and enable extrapolation of sensing likelihood function $P(\hat{\mathbf{z}}|\mathbf{x})$ from the communication channel knowledge distribution $P(\mathbf{w}|\mathbf{q})$, regardless of its specific RCS or scattering characteristics.

Specifically, assuming independence of the channel knowledge of the $L$ paths, the likelihood function $ P(\hat{\mathbf{z}}|\mathbf{x}) $ is expressed as
\begin{equation}
	\begin{aligned}
		P(\hat{\mathbf{z}}|\mathbf{x}) =\prod_{l=1}^{L}P(\hat{\mathbf{z}}_l|\mathbf{x}).
	\end{aligned}
\label{eq32}
\end{equation}
Then, the sensing likelihood function for each path can be written as
\begin{equation}
	\begin{aligned}
		P(\hat{\mathbf{z}}_l|\mathbf{x})=&P(\phi_l, \theta_l, \tau_l|\mathbf{x}) \\
		=&P(\theta_{l_T}, \theta_{l_R}, \tau_{l_R} + \tau_{l_T}|\mathbf{q}=\mathbf{x})\\
		=&\int\int P(\theta_{l_T}, \tau_{l_T}|\mathbf{q}=\mathbf{x}) P(\theta_{l_R},\tau_{l_R}|\mathbf{q}=\mathbf{x})\\
		&\times\delta(\tau_l-(\tau_{l_T}+\tau_{l_R})) d\tau_{l_T}d\tau_{l_R},
	\end{aligned}
	\label{likelihood3}
\end{equation}
where the second equation is due to the equations in (\ref{equa}), and the forward and reverse path indices $ (l_T,l_R) $ are determined from the composite path index $ l $ via the mapping $l_T=\lfloor(l-1)/L'\rfloor+1 $ and $ l_R=(l-1)(mod \ L')+1 $.
Besides, the Dirac delta function $\delta(\tau_l - (\tau_{l_T} + \tau_{l_R}))$ enforces the delay relationship \(\tau_l = \tau_{l_R} + \tau_{l_T}\).

The formula (\ref{likelihood3}) captures the combined effect of downlink (forward) and uplink (reverse) paths, enabling inference of the sensing likelihood function $ P(\hat{\mathbf{z}}|\mathbf{x}) $ from the communication channel knowledge distribution 	$ P_{\mathcal{M}}(\mathbf{w}|\mathbf{q}) $ at the target location $\mathbf{x}$, as elaborated in the following.

\subsection{CADM-enabled sensing}

This subsection details the specific implementation of environment-aware sensing using CKM, with a focus on a specific type of CKM called CADM. CADM leverages historical multipath angle and delay at various UE locations $\mathbf{q}$ within the environment $E$ to learn the location-specific statistical distribution of angular and temporal characteristics. This mapping is mathematically expressed as
\begin{equation}
	\mathcal{M}: \mathbf{q} \rightarrow P_{\mathcal{M}}(\mathbf{w}|\mathbf{q}),
\end{equation}
where
\begin{equation}
	\mathbf{w}=[\mathbf{w}_1^T,\mathbf{w}_2^T,...,  \mathbf{w}_{l'}^T,...,\mathbf{w}_{L'}^T]^T\in \mathbb{R}^{2L'\times1},
\end{equation}
\begin{equation}
	\mathbf{w}_{l'}=[\theta_{l'},  \tau_{l'}].
\end{equation}
Here, the channel knowledge vector $ \mathbf{w}_{l'} $ is defined with only AoA $ \theta_{l'} $ and delay $ \tau_{l'} $.

CADM can be mathematically represented in various forms. For instance, in an idealized deterministic case where path parameters are known perfectly, the CADM could be modeled as a series of impulse functions corresponding to the discrete AoA and delay values (e.g., $\sum_{l'} \delta(\mathbf{w} - \mathbf{w}_{l'})$).
However, a more practical approach is to use a probabilistic representation. For convenience, this paper adopts a Gaussian model to represent the location-specific angle-delay channel knowledge $P_{\mathcal{M}}(\mathbf{w}|\mathbf{q})$, which can be written as

\begin{equation}
\small
	\begin{aligned}
		\ln \ &P_{\mathcal{M}}(\mathbf{w}|\mathbf{q}) \\
		=& \sum_{l'=1}^{L'} \ln P_{\mathcal{M}}(\mathbf{w}_{l'}|\mathbf{q})\\
		=&\sum_{l'=1}^{L'}\ln\big(\mathcal{N}(\theta_{l'}; \mu_{\theta,{l'}}(\mathbf{q}), \sigma_{\theta,{l'}}^2(\mathbf{q}))\\
		&\times\mathcal{N}(\tau_{l'}; \mu_{\tau,{l'}}(\mathbf{q}), \sigma_{\tau,{l'}}^2(\mathbf{q}))\big)
	\end{aligned}
	\label{cadmd}
\end{equation}
where $\mu_{\theta,l'}(\mathbf{q})$ and $\sigma_{\theta,l'}^2(\mathbf{q})$ are the mean and variance of the AoA $\theta_{l'}$, and $\mu_{\tau,l'}(\mathbf{q})$ and $\sigma_{\tau,l'}^2(\mathbf{q})$ are the mean and variance of the delay $\tau_{l'}$ for the $l'$th path at UE location $\mathbf{q}$. These parameters are location-dependent, reflecting the unique channel characteristics at each $\mathbf{q}$ in the environment. The number of paths $L'$ is chosen to capture the dominant propagation paths, with more complex environments requiring a larger $L'$.
The location-specific probabilistic modeling in (\ref{cadmd}) is crucial for the framework's effectiveness in dynamic environments. While dominant propagation paths are determined by large and fixed structures, the movement of non-target objects (e.g., other vehicles or pedestrians) introduces statistical fluctuations. The learned variance captures these variations from historical data; a path that is frequently perturbed by moving scatterers will exhibit a larger spread in the training data, for which the model learns a higher variance. In this way, the typical environment dynamics are already embedded within the probabilistic CKM. The framework thus assumes a statistically stable environment, rather than a perfectly static one. This implies an environment where large-scale features (e.g., buildings, roads) are stable, while the variance learned by CKM accounts for the statistical fluctuations caused by dynamic elements such as vehicles and pedestrians. This assumption aligns well with typical urban environments.

To model the dependence of these parameters on $\mathbf{q}$, CADM employs an FCNN to learn the mapping from the UE location $ \mathbf{q} $ to the mean and variance parameters in the Gaussian model in (\ref{cadmd}). For example, if a two-layer FCNN is used with the input being the UE location $\mathbf{q} \in \mathbb{R}^2$, the forward process is defined as
\begin{equation}
\small
	\begin{aligned}
		&\mathbf{h}_1 = \sigma(\mathbf{W}_1 \mathbf{q} + \mathbf{b}_1), \\
		&\{ \mu_{\theta,{l'}}(\mathbf{q}),\mu_{\tau,{l'}}(\mathbf{q}), \sigma_{\theta,{l'}}^2(\mathbf{q}), \sigma_{\tau,{l'}}^2(\mathbf{q})\}_{{l'}=1}^{L'} 
		= \mathbf{W}_2 \mathbf{h}_1 + \mathbf{b}_2,
	\end{aligned}
\label{fcnn}
\end{equation}
where $\mathbf{W}_1$, $\mathbf{b}_1$, $\mathbf{W}_2$, and $\mathbf{b}_2$ are the weight matrices and bias vectors of the FCNN, and $\sigma(\cdot)$ is the activation function (e.g., ReLU).
In communication systems, CADM utilizes the inferred angle and delay distributions based on UE locations $\mathbf{q}$ to support tasks such as channel estimation and delay alignment modulation (DAM) \cite{lu2023delay}. 
The novel contribution of this paper lies in extending CADM to sensing, where the target is treated as a virtual UE reflecting signals back to the BS. 

To derive the sensing likelihood function $P(\hat{\mathbf{z}}|\mathbf{x})$ at the target location $ \mathbf{x} $, the CADM is adapted for the sensing task. This adaptation involves two key steps. First, since the target is treated as a virtual UE, the CADM's location input $ \mathbf{q} $ is set to the target location $ \mathbf{x} $. Second, the specific Gaussian distribution for the channel knowledge provided by CADM at this location, as defined in (\ref{cadmd}), is substituted into the general form of the likelihood function in (\ref{likelihood3}) as

\begin{equation}
\small
	\begin{aligned}
		P(&\theta_{l_T}, \tau_{l_T}|\mathbf{q}=\mathbf{x})\\
		&=P_\mathcal{M}(\mathbf{w}_{l'}=(\theta_{l_T}, \tau_{l_T})|\mathbf{q}=\mathbf{x})\\
		&=\mathcal{N}(\theta_{l_T}; \mu_{\theta,{l_T}}(\mathbf{x}), \sigma_{\theta,{l_T}}^2(\mathbf{x}))\\
		&\quad\times\mathcal{N}(\tau_{l_T}; \mu_{\tau,{l_T}}(\mathbf{x}), \sigma_{\tau,{l_T}}^2(\mathbf{x}))
	\end{aligned}
	\label{eq-forward-path}
\end{equation}
and 
\begin{equation}
\small
	\begin{aligned}
		P(&\theta_{l_R}, \tau_{l_R}|\mathbf{q}=\mathbf{x})\\
		&=P_\mathcal{M}(\mathbf{w}_{l'}=(\theta_{l_R}, \tau_{l_R})|\mathbf{q}=\mathbf{x})\\
		&=\mathcal{N}(\theta_{l_R}; \mu_{\theta,{l_R}}(\mathbf{x}), \sigma_{\theta,{l_R}}^2(\mathbf{x}))\\
		&\quad\times\mathcal{N}(\tau_{l_R}; \mu_{\tau,{l_R}}(\mathbf{x}), \sigma_{\tau,{l_R}}^2(\mathbf{x})).
	\end{aligned}
	\label{eq-reverse-path}
\end{equation}
Then, the log-likelihood of the sensing channel knowledge can be written as (\ref{CADMlike}) shown on the top of the next page, where the second equation is due to (\ref{likelihood3}).
In (\ref{eq-forward-path}) and (\ref{eq-reverse-path}), we assume that for a given path $l'$, the angle $\theta_{l'}$ and delay $\tau_{l'}$ are statistically independent. This assumption is based on the underlying physics of multipath propagation in complex environments. From a physical standpoint, the parameters are determined by different geometric properties: The AoA is primarily dictated by the geometric orientation of the final scatterer relative to the receiver array. The path delay, in contrast, is a function of the entire path length traveled by the signal, which may involve multiple different scattering objects before reaching that final scatterer. In the NLoS multipath environment, the factors contributing to the total path length (and thus delay) are numerous and distinct from the geometric angle of the final hop. Therefore, in complex environments, the statistical relationship between the total path length and the final arrival angle is generally weak. It is also noted that the CKM framework is not fundamentally restricted by this assumption. The model could be extended to a full multivariate Gaussian distribution.

\begin{figure*}[!ht]
	\begin{equation}
	\small
		\begin{aligned}
			\ln P(\hat{\mathbf{z}}|\mathbf{x})=&\sum_{l=1}^{L} \ln P(\mathbf{z}_l|\mathbf{x})=\sum_{l=1}^{L} \ln P(\phi_l, \theta_l, \tau_l|\mathbf{x})\\
			=&\sum_{l=1}^{L}\ln \left(\int\int P(\theta_{l_T}, \tau_{l_T}|\mathbf{q}=\mathbf{x}) P(\theta_{l_R},\tau_{l_R}|\mathbf{q}=\mathbf{x})\delta(\tau_l-(\tau_{l_T}+\tau_{l_R})) d\tau_{l_T}d\tau_{l_R}\right)\\
			=&\sum_{l=1}^{L}\ln \left(\int\int\mathcal{N}(\theta_{l_T}; \mu_{\theta,{l_T}}(\mathbf{x}), \sigma_{\theta,{l_T}}^2(\mathbf{x}))\mathcal{N}(\tau_{l_T}; \mu_{\tau,{l_T}}(\mathbf{x}), \sigma_{\tau,{l_T}}^2(\mathbf{x}))
			\mathcal{N}(\theta_{l_R}; \mu_{\theta,{l_R}}(\mathbf{x}), \sigma_{\theta,{l_R}}^2(\mathbf{x}))\right.\\
			&\times\mathcal{N}(\tau_{l_R}; \mu_{\tau,{l_R}}(\mathbf{x}), \sigma_{\tau,{l_R}}^2(\mathbf{x}))\delta(\tau_l-(\tau_{l_T}+\tau_{l_R})) d\tau_{l_T}d\tau_{l_R}\Big)\\
			=&\sum_{l=1}^{L}\ln\Big(\mathcal{N}\left(\phi_{l}; \mu_{\theta,{l_T}}\left(\mathbf{x}\right), \sigma_{\theta,{l_T}}^2\left(\mathbf{x}\right)\right)
			\mathcal{N}\left(\theta_{l}; \mu_{\theta,{l_R}}\left(\mathbf{x}\right), \sigma_{\theta,{l_R}}^2\left(\mathbf{x}\right)\right)\\
			&\times\mathcal{N}\left(\tau_{l}; \mu_{\tau,{l_T}}\left(\mathbf{x}\right)+\mu_{\tau,{l_R}}\left(\mathbf{x}\right), \sigma_{\tau,{l_T}}^2\left(\mathbf{x}\right)+\sigma_{\tau,{l_R}}^2\left(\mathbf{x}\right)\right)\Big).
		\end{aligned}
		\label{CADMlike}
	\end{equation}
	\hrulefill
\end{figure*}

\begin{figure*}[ht]
	\begin{equation}
	\small
		\begin{aligned}
			\nabla_\mathbf{x}(-\ln P(\hat{\mathbf{z}}|\mathbf{x}))
			=&\nabla_\mathbf{x}\sum_{l=1}^{L}-\ln \left(\mathcal{N}(\phi_{l}; \mu_{\theta,{l_T}}(\mathbf{x}), \sigma_{\theta,{l_T}}^2(\mathbf{x}))
			\mathcal{N}(\theta_{l}; \mu_{\theta,{l_R}}(\mathbf{x}), \sigma_{\theta,{l_R}}^2(\mathbf{x}))\right.\\
			&\times\mathcal{N}(\tau_{l}; \mu_{\tau,{l_T}}(\mathbf{x})+\mu_{\tau,{l_R}}(\mathbf{x}), \sigma_{\tau,{l_T}}^2(\mathbf{x})+\sigma_{\tau,{l_R}}^2(\mathbf{x}))\big)\\
			=&-\sum_{l=1}^{L}\left( \frac{\phi_l-\mu_{\theta,{l_T}}(\mathbf{x})}{\sigma_{\theta,{l_T}}^2(\mathbf{x})}\nabla_\mathbf{x}\mu_{\theta,{l_T}}(\mathbf{x})+\frac{\left(\phi_l-\mu_{\theta,{l_T}}(\mathbf{x})\right)^2-\sigma_{\theta,{l_T}}^2(\mathbf{x})}{2\sigma_{\theta,{l_T}}^4(\mathbf{x})}\nabla_\mathbf{x}\sigma_{\theta,{l_T}}^2(\mathbf{x})\right.\\
			&+\frac{\theta_l-\mu_{\theta,{l_R}}(\mathbf{x})}{\sigma_{\theta,{l_R}}^2(\mathbf{x})}\nabla_\mathbf{x}\mu_{\theta,{l_R}}(\mathbf{x})+\frac{\left(\theta_l-\mu_{\theta,{l_R}}(\mathbf{x})\right)^2-\sigma_{\theta,{l_R}}^2(\mathbf{x})}{2\sigma_{\theta,{l_R}}^4(\mathbf{x})}\nabla_\mathbf{x}\sigma_{\theta,{l_R}}^2(\mathbf{x})\\
			&+\frac{\tau_l-\mu_{\tau,{l_T}}(\mathbf{x})-\mu_{\tau,{l_R}}(\mathbf{x})}{\sigma_{\tau,{l_T}}^2(\mathbf{x})+\sigma_{\tau,{l_R}}^2(\mathbf{x})}\nabla_\mathbf{x}\big(\mu_{\tau,{l_T}}(\mathbf{x})+\mu_{\tau,{l_R}}(\mathbf{x})\big)\\
			&+\frac{\left(\tau_l-\mu_{\tau,{l_T}}(\mathbf{x})-\mu_{\tau,{l_R}}(\mathbf{x})\right)^2-\sigma_{\tau,{l_T}}^2(\mathbf{x})-\sigma_{\tau,{l_R}}^2(\mathbf{x})}{2(\sigma_{\tau,{l_T}}^2(\mathbf{x})+\sigma_{\tau,{l_R}}^2(\mathbf{x}))^2}\nabla_\mathbf{x}\big(\sigma_{\tau,{l_T}}^2(\mathbf{x})+\sigma_{\tau,{l_R}}^2(\mathbf{x})\big)\bigg).
		\end{aligned}
		\label{daoshu}
	\end{equation}
	\hrulefill
\end{figure*}

To estimate the sensing target location $\mathbf{x}$ using gradient descent as specified in (\ref{xk}), the gradient of $\ln P(\hat{\mathbf{z}}|\mathbf{x})$ with respect to $\mathbf{x}$ is required. This gradient is given by (\ref{daoshu}) shown on the top of the next page, where the gradients $\nabla_{\mathbf{x}}\mu_{\theta,l_T}(\mathbf{x})$, $\nabla_{\mathbf{x}}\sigma_{\theta,l_T}^2(\mathbf{x})$, $\nabla_{\mathbf{x}}\mu_{\tau,l_T}(\mathbf{x})$, $\nabla_{\mathbf{x}}\sigma_{\tau,l_T}^2(\mathbf{x})$, $\nabla_{\mathbf{x}}\mu_{\theta,l_R}(\mathbf{x})$, $\nabla_{\mathbf{x}}\sigma_{\theta,l_R}^2(\mathbf{x})$, $\nabla_{\mathbf{x}}\mu_{\tau,l_R}(\mathbf{x})$, and $\nabla_{\mathbf{x}}\sigma_{\tau,l_R}^2(\mathbf{x})$ can be obtained based on the structure of FCNN as (\ref{fcnn}) by substituting $ \mathbf{q} $ with $ \mathbf{x} $. For example, the gradient of a parameter such as $\mu_{\theta,l_T}(\mathbf{x})$ is written as 
\begin{equation}
	\nabla_{\mathbf{x}} \mu_{\theta,l_T}(\mathbf{x}) = \mathbf{a}^T \mathbf{W}_2 \cdot \sigma'(\mathbf{W}_1 \mathbf{x} + \mathbf{b}_1) \cdot \mathbf{W}_1,
\end{equation}
where $\mathbf{a}\in \mathbb{R}^{4L'\times 1}$ is a selection vector with a $ 1 $ at the position corresponding to $\mu_{\theta,l_T}(\mathbf{x})$ and $ 0 $ elsewhere, $\mathbf{W}_1$, $\mathbf{W}_2$, $\mathbf{b}_1$ are FCNN weights and biases, and $\sigma'(\cdot)$ is the derivative of the activation function (such as ReLU). These closed-form expressions enable computation of $\nabla_{\mathbf{x}} (-\ln P(\hat{\mathbf{z}}|\mathbf{x}))$, facilitating optimization of $\mathbf{x}$ via (\ref{xk}).
Alternatively, these gradients can be computed efficiently using automatic differentiation in PyTorch by enabling gradient tracking for the location input, calling $ backward() $ on the output, and then accessing the input's $ .grad $ attribute. Besides, the $ torch.autograd.grad() $ function provides a more direct alternative for this computation.

\looseness=-1 In summary, to estimate the location $\mathbf{x}$ of the sensing target in the challenging NLoS environment, parameters such as AoD, AoA, and delay for each path are first estimated 
based on the echo signal $\mathbf{y}(t)$ in (\ref{eqm}) using well-established algorithms (e.g., MUSIC, ESPRIT, or compressed sensing \cite{dai2025tutorial}), forming the observation vector $\hat{\mathbf{z}}$.
The methods cited, such as MUSIC or ESPRIT, are mentioned only as well-established, high-resolution examples, not as mandatory components. For dynamic scenarios involving high-speed targets, where low latency and real-time capability are required, our framework can be seamlessly integrated with various estimation algorithms (e.g., FFT-based) specifically designed for high-mobility tracking. The core contribution of this work lies in intelligently utilizing the outputs $\hat{\mathbf{z}}$ of any such algorithms for NLoS localization. Then, substituting $\hat{\mathbf{z}}$ into (\ref{daoshu}) allows computation of the gradient, enabling estimation of the target location $\mathbf{x}$ via gradient descent in (\ref{xk}). Note that while the algorithm is presented for the most challenging NLoS scenarios, it is applicable to the LoS case as well. The pseudo-code of CKM-enabled environment-aware sensing is elaborated in Algorithm~1.

\begin{algorithm}[htbp]
	\caption{CKM-Enabled LoS and NLoS Sensing}
	\begin{algorithmic}[1]
		\Require Received echoes $\mathbf{y}(t)$, CADM $P_{\mathcal{M}}(\mathbf{w}|\mathbf{q})$ represented using an FCNN such as (\ref{fcnn}), learning rate $\eta$, initial estimate of the sensing target location $\mathbf{x}^{(0)}$
		\Ensure Estimated sensing target location $\hat{\mathbf{x}}$
		\State Estimate the angle and delay information from $\mathbf{y}(t)$ using various parameter estimation algorithms (such as MUSIC, ESPRIT), and denote the estimated vector as $\hat{\mathbf{z}}$
		\State Initialize $k = 0$
		\While{not converged}
		\State {\small Compute the gradient $\nabla_{\mathbf{x}} (-\ln P(\hat{\mathbf{z}}|\mathbf{x})) \big|_{\mathbf{x}^{(k)}}$ based on (\ref{daoshu})}
		\State  {\small Update sensing target location estimation:} 
		\Statex \hspace*{2em} {\small$\mathbf{x}^{(k+1)} = \mathbf{x}^{(k)} - \eta \nabla_{\mathbf{x}} (-\ln P(\hat{\mathbf{z}}|\mathbf{x})) \big|_{\mathbf{x}^{(k)}}$}
		\State $k \gets k + 1$
		\EndWhile
		\State \textbf{Return} $\hat{\mathbf{x}} = \mathbf{x}^{(k)}$
	\end{algorithmic}
\end{algorithm}

\section{Simulation Results}
In this section, we present the simulation results to evaluate the performance of the proposed CKM-based sensing framework. We first describe the simulation setup, followed by the training process of the FCNN model used for channel knowledge prediction in (\ref{fcnn}). Then, we analyze the sensing performance with several benchmarks in both pure NLoS and mixed LoS and NLoS scenarios. The proposed method is compared with the classic geometry-based methods. Finally, we investigate the CRLB as a function of angle and delay estimation errors to further validate the robustness of our approach.

\begin{figure}[htbp]
	\centering{\includegraphics[width=.43\textwidth]{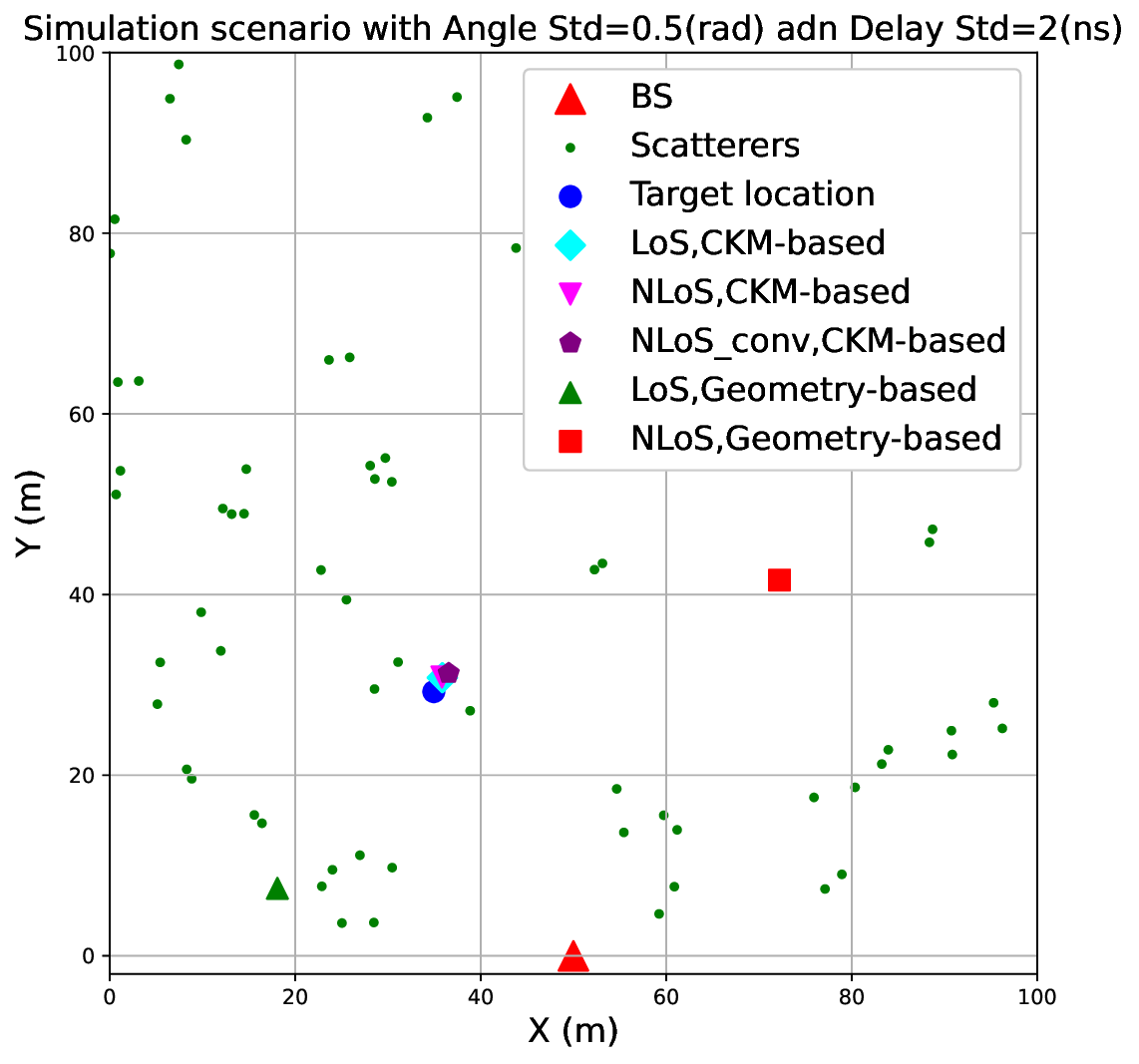}}  	
	\caption{Simulation setup for ISAC system and sensing results of different methods.} 
	\label{R1}  
\end{figure}

The simulation is conducted in a 100 m $\times$ 100 m square area, where a BS is located at $ (50, 0) m $. Besides, there are several scatterers randomly distributed within the area, and one target is placed at a random location for each sensing process. 
To model the statistical nature of a dynamic environment, we incorporated small random location variations for the scatterers during the data generation process. This ensures that the training data reflects statistical fluctuations, allowing the FCNN to learn a meaningful location-dependent mean and variance for the channel parameters, rather than just a fixed value.
The locations of the BS, scatterers, and target are illustrated in Fig. \ref{R1}, where the BS is marked with a red triangle, scatterers with green circles, and the sensing target with a blue circle. Two sensing scenarios are considered:
\begin{itemize}
	\item \textbf{Mixed LoS/NLoS}: The direct LoS path between BS and target is retained, alongside NLoS paths from scatterers.
	\item \textbf{Pure NLoS}: The LoS path between the BS and target is removed, simulating LoS blocked settings, with sensing relying only on NLoS paths from scatterers.
\end{itemize}

\begin{figure}[htbp]
	\centering{\includegraphics[width=.40\textwidth]{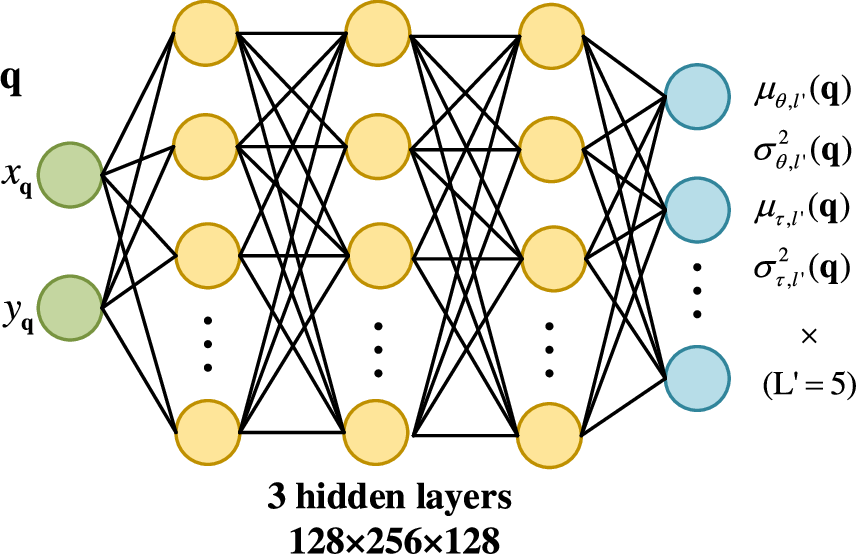}}  	
	\caption{The structure of FCNN.} 
	\label{FCNN}  
\end{figure}

To construct the CKM, we use an FCNN as shown in Fig. \ref{FCNN}. It consists of five layers: an input layer with 2 neurons (corresponding to the 2D coordinates of UE location $\mathbf{q}$), three hidden layers with $ 128\times256\times128 $ neurons using LeakyReLU activation with a negative slope of 0.01, and an output layer with $4L'$ neurons, where $L'=5$ is the number of Gaussian components in (\ref{cadmd}). The output includes the angle mean $\mu_{\theta,l'}$, the angle variance $\sigma_{\theta,l'}^2$, the delay mean $\mu_{\tau,l'}$, and the delay variance $\sigma_{\tau,l'}^2$ for each path $l'$.
Besides, the training dataset is generated by sampling 10,000 points uniformly within the 100 m $\times$ 100 m area. This dataset was empirically found to be sufficient for the FCNN to converge and was partitioned into a training set (80\%) and a validation set (20\%).
Note that in practical deployments, it is unnecessary to physically collect all 10,000 samples. As mentioned in Section~I, advanced CKM construction methods based on generative AI \cite{fu2025ckmdiff,jin2024i2i} can complete a large-scale dataset from only a small fraction of real-world measurements, significantly reducing the data collection burden.
For each sampled UE location in the training dataset, the groundtruth channel is generated by considering all possible single-bounce paths via the scatterers. The delay for each path is calculated based on its total geometric length, which is the distance from the BS to the scatterer and then to the UE. The complex gain of each path is determined by two main factors: the path-length-dependent attenuation and a unique random reflection coefficient assigned to each scatterer.
Although all paths are generated, the CADM is designed to learn the $ L'=5 $ dominant ones. Therefore, a path selection mechanism is applied: the 5 paths with the highest power, as determined by their path gains, are chosen. The angles and delays of these 5 selected paths constitute the groundtruth data used for training the FCNN, and the training loss is computed as the mean squared error (MSE) between the predicted and true channel parameters.
The network was trained for 500 epochs using the Adam optimizer, with an initial learning rate of 0.01 and a batch size of 128.

An important practical consideration for the proposed sensing algorithm (Algorithm~1) is the initialization of the target location $\mathbf{x}^{(0)}$. Since the sensing likelihood function $P(\hat{\mathbf{z}}|\mathbf{x})$ in (\ref{CADMlike}) can be non-convex, the gradient-descent-based optimization may be sensitive to this initial value and could converge to a local optimum. To mitigate this sensitivity in our simulations and increase the probability of converging to the global optimum, we adopt a multi-start strategy. Specifically, ten initial points are sampled uniformly from the $100~m \times 100~m$ sensing area, and Algorithm~1 is run for each. The final location estimate is then selected as the one that yields the highest likelihood value $P(\hat{\mathbf{z}}|\mathbf{x})$ after convergence.
While this multi-start random initialization is used for our evaluation, other sophisticated strategies could be employed in practical applications. For instance, a rough initial guess can be obtained by using the parameters (e.g., shortest delay) of the strongest received path in a geometric localization formula, similar to (\ref{xyco}). Another effective method is a coarse grid search, where the likelihood $P(\hat{\mathbf{z}}|\mathbf{x})$ is evaluated at predefined grid points, and the point with the maximum likelihood serves as a high-quality starting value for the subsequent optimization.

The performance of the proposed CKM-based sensing framework is compared against the following benchmarks in LoS and NLoS scenarios:
\begin{itemize}
	\item \textbf{LoS, Geometry-based:} Uses delay and angle of LoS path for geometric target sensing as (\ref{xyco}).
	\item \textbf{NLoS, Geometry-based:} Assumes the shortest-delay path as LoS for geometric localization.
	\item \textbf{LoS, CKM-based:} Employs CKM in LoS scenarios, leveraging both LoS and NLoS paths for accurate localization, modeling up to $L'^2$ non-reciprocal paths for sensing as (\ref{eqm}).
	\item \textbf{NLoS, CKM-based:} Employs CKM in NLoS scenarios, leveraging only NLoS paths for localization of the sensing target, modeling up to $L'^2$ non-reciprocal paths for sensing as (\ref{eqm}).
	\item \textbf{NLoS\_conv, CKM-based:} Applies CKM in NLoS scenarios but with a conventional sensing model ($L'$ reciprocal paths) (\ref{eqmonosignaltra}), rather than $L'^2$ paths as with the proposed sensing model in (\ref{eqm}).
\end{itemize}

\begin{figure}[htbp]
	\centering{\includegraphics[width=.43\textwidth]{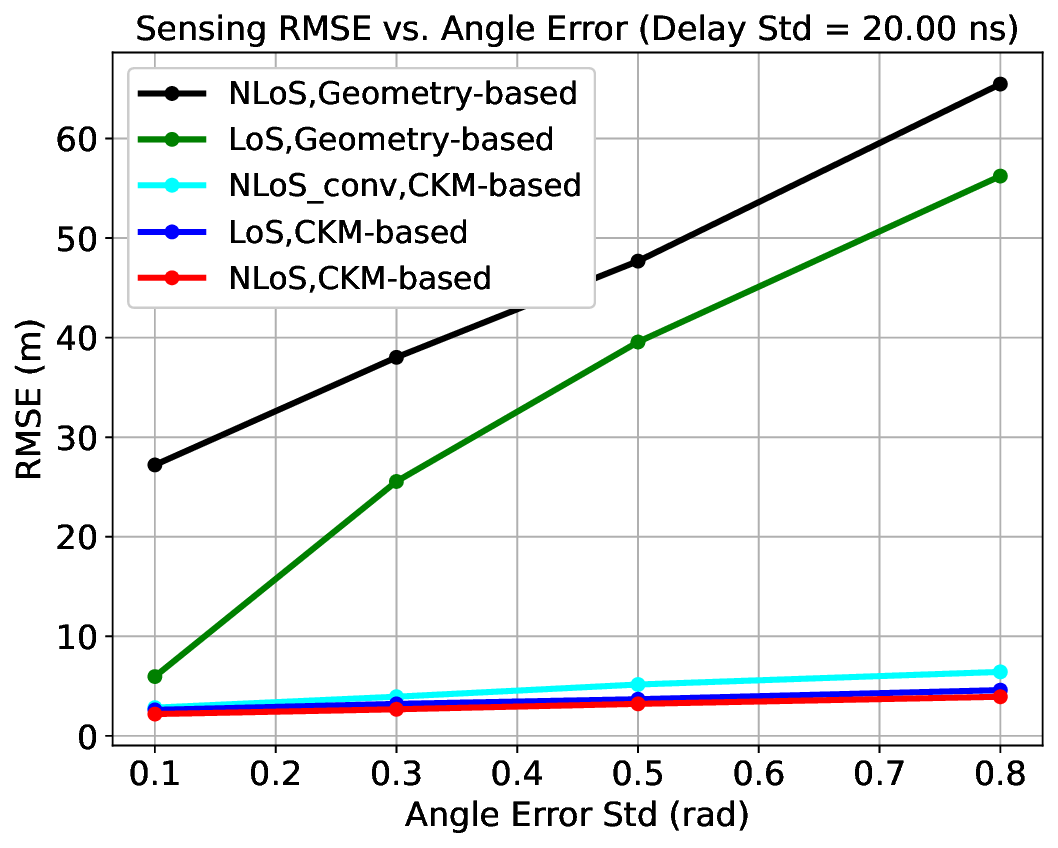}}  	
	\caption{Sensing RMSE versus angle error.} 
	\label{MSEangle}  
\end{figure}

\begin{figure}[htbp]
	\centering{\includegraphics[width=.43\textwidth]{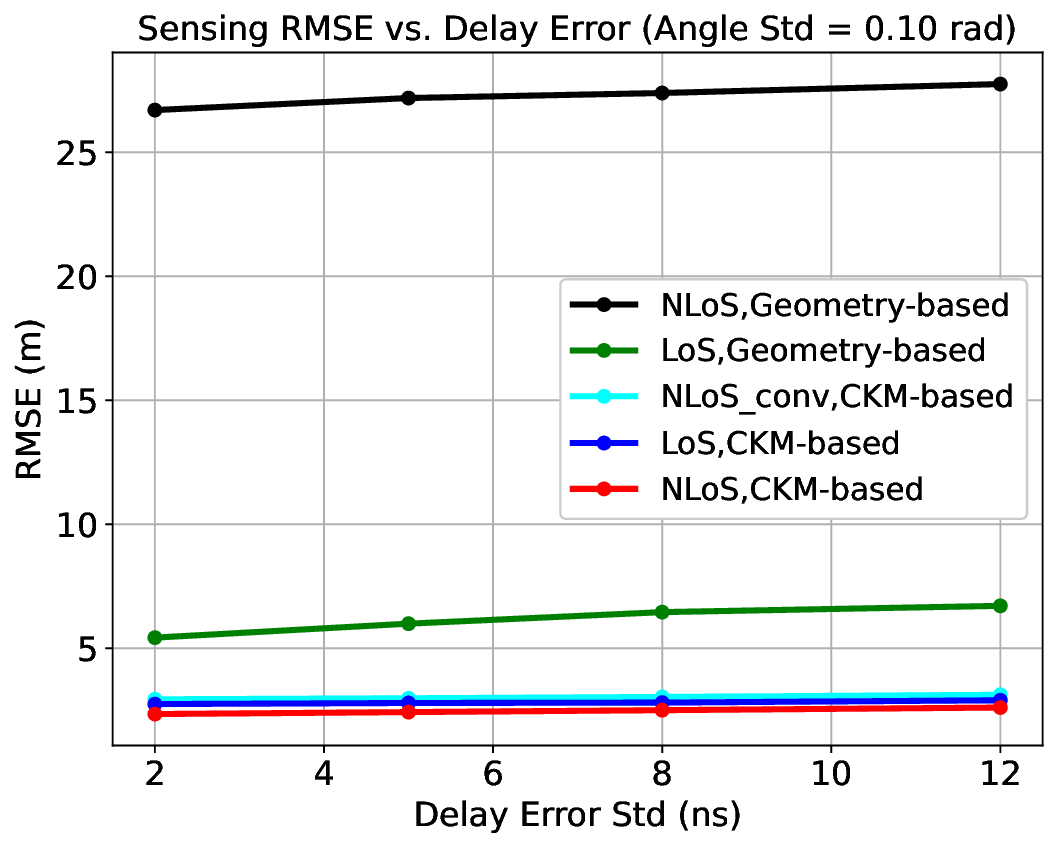}}  	
	\caption{Sensing RMSE versus delay error.} 
	\label{MSEdelay}  
\end{figure}

Fig. \ref{MSEangle} presents the RMSE for the location estimation of the sensing target with different methods vs. angle estimation error $ \sigma_\theta $ and $ \sigma_\phi $, with fixed delay std $ \sigma_\tau= 20 $ ns.
It can be seen that the CKM-based methods consistently demonstrate significantly lower RMSE than geometry-based approaches across all angle error levels, by orders of magnitude. Such dramatic performance gains are attributed to the utilization of CADM, which exploits a rich set of NLoS path information, including AoA, AoD and delay.
In contrast, even in the mixed LoS/NLoS scenarios, the geometry-based method relies solely on the LoS path’s angle and delay, where angle errors severely degrade performance due to their tight coupling with location as (\ref{conmoe}). Even with accurate delay estimates, RMSE increases sharply as angle error grows.
For the challenging NLoS scenario, the geometry-based method fails to give any meaningful estimation for the sensing target location, since it erroneously assumes the shortest-delay NLoS path as LoS, leading to substantial localization errors.
CKM-based methods, by leveraging multiple paths, decouple angle and delay contributions, resulting in much slower RMSE degradation with increasing angle error, as precise delay measurements mitigate the impact of angle inaccuracies. In the pure NLoS scenario, the CKM-based method, employing the proposed generalized sensing model as (\ref{eqm}) with up to $L'^2$ non-reciprocal paths, outperforms CKM-based NLoS\_conv, which uses a conventional model as (\ref{eqmonosignaltra}) with only $L'$ reciprocal paths. The reduced path diversity in the conventional model limits its information, yielding higher RMSE. Surprisingly, CKM-based NLoS achieves even lower RMSE than CKM-based LoS, especially when the angle errors are large. This may seem counterintuitive at first glance, but is expected because NLoS paths, with distinct AoA, AoD and delay, provide three parameters per path, compared to two for LoS paths (AoA = AoD, delay). The increased path count ($L'^2$ vs. $L'$) and richer information per path in NLoS scenarios enhance robustness to angle errors when properly exploited through CKM, enabling superior localization accuracy despite the absence of a LoS path. Similar observations can be obtained from Fig.~\ref{MSEdelay}, which plots the sensing RMSE as a function of the delay estimation error $ \sigma_\tau $.

To further illustrate the sensing performance, Fig. \ref{R1} presents a specific example with an angle standard deviation of $ \sigma_{\phi}=\sigma_{\theta}=0.5 $ rad and a delay standard deviation of $ \sigma_{\tau}=2 $ ns. The groundtruth target location is marked with a blue circle, while the estimated locations from the geometry-based LoS, geometry-based NLoS, CKM-based NLoS, CKM-based LoS and CKM-based NLoS\_conv methods, are marked with a green triangle, a red square, a cyan diamond, a magenta downward triangle, and a purple pentagon, respectively. The CKM-based estimations are concentrated around the groundtruth location, whereas the two geometry-based estimations deviate significantly due to the angle and delay estimation error or incorrect assumption of the LoS path.

\begin{figure}[htbp]
	\centering{\includegraphics[width=.45\textwidth]{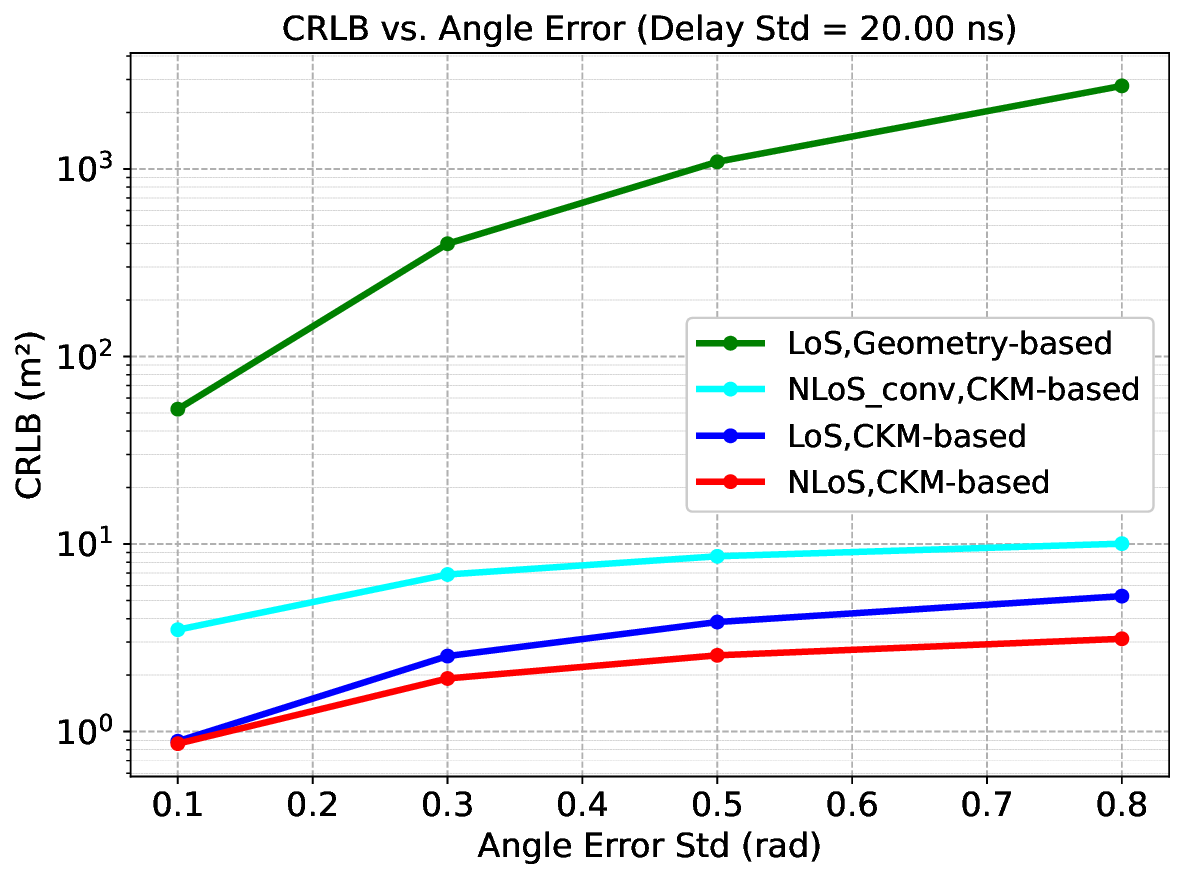}}  	
	\caption{CRLB versus angle error.} 
	\label{CRLBangle}  
\end{figure}

\begin{figure}[htbp]
	\centering{\includegraphics[width=.45\textwidth]{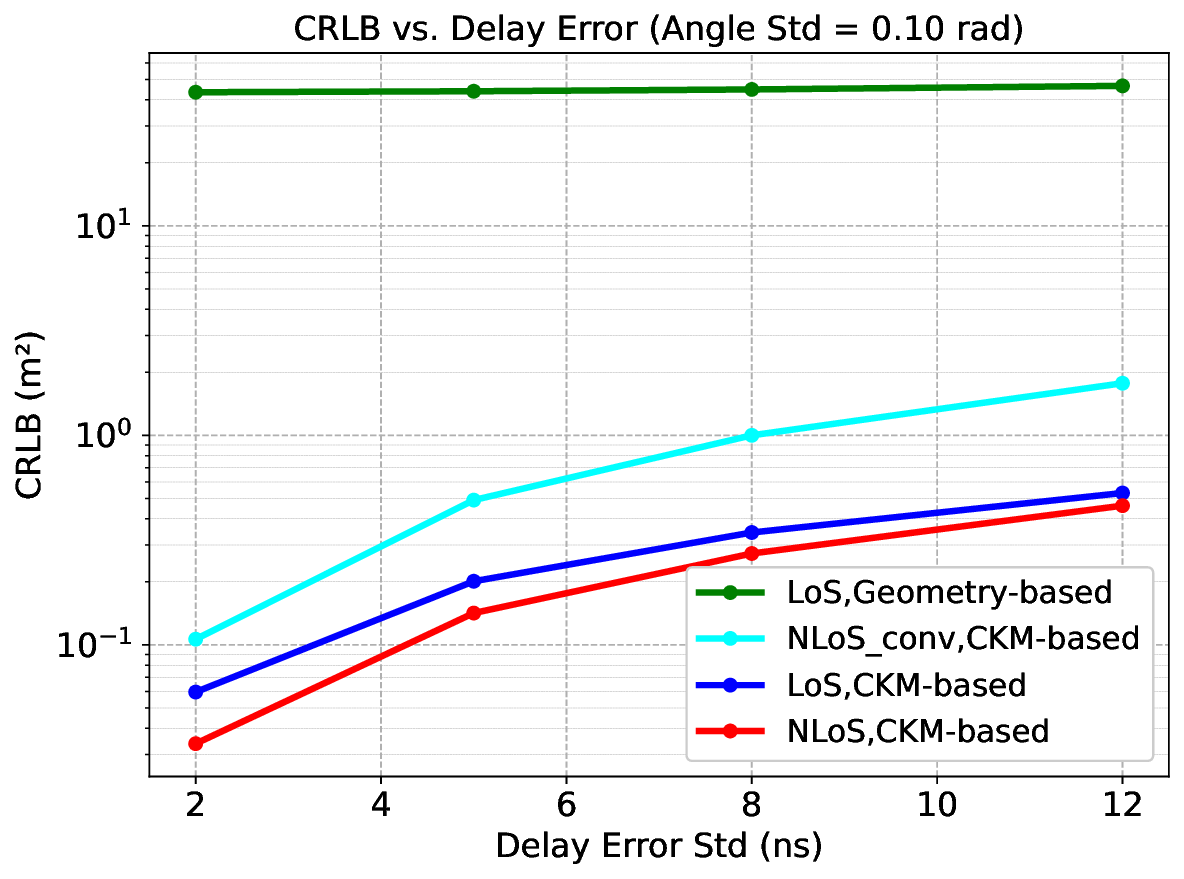}}  	
	\caption{CRLB versus delay error.} 
	\label{CRLBdelay}  
\end{figure}

Following the MSE analysis, the theoretical performance bounds are explored through the CRLB to highlight localization performance limits. CRLB is analyzed vs. angle error (with delay error std $\sigma_\tau = 20$ ns) and delay error $\sigma_\tau$ (with angle error std $\sigma_\theta=\sigma_\phi = 0.10$ rad) as shown in Fig. \ref{CRLBangle} and \ref{CRLBdelay}, respectively. The geometry-based NLoS is excluded, as unknown scatterer location prevents CRLB computation. 
For the proposed CKM-based methods, the CRLB is computed as the inverse of the FIM, and the FIM is calculated based on the first equation in (\ref{fishermatrix}), where the required gradient of the log-likelihood $ \nabla_\mathbf{x}(-\ln P(\hat{\mathbf{z}}|\mathbf{x})) $ is provided in (\ref{daoshu}) and is computed efficiently via automatic differentiation.
The geometry-based LoS, CKM-based LoS, NLoS, and NLoS\_conv are evaluated, with CRLB differences due to different path utilization for sensing and distinct sensing functions mapping the relationship between target location and channel knowledge.

Fig. \ref{CRLBangle} shows the CRLB as a function of angle estimation error $ \sigma_\theta=\sigma_\phi $. 
The geometry-based LoS method exhibits a severe CRLB increase as angle errors grow. 
This is due to its dependence on a single LoS path, where the sensing function tightly couples angle and delay errors, as shown in (\ref{conmoe}), indicating the method’s vulnerability to measurement inaccuracies. In contrast, all CKM-based methods demonstrate consistently lower CRLBs with a mild increase. Their superior performance stems from sensing functions that incorporate a broader set of NLoS paths and decouple angle and delay errors, allowing precise delay measurements to compensate for angle inaccuracies and thus enhancing localization robustness.
Among the CKM-based methods, the proposed NLoS model (using $L'^2$ non-reciprocal paths) and the LoS model both outperform the conventional NLoS\_conv model (using $L'$ reciprocal paths), as the latter retains less path information. Notably, at higher $\sigma_\theta$, the CKM-based NLoS method achieves a lower CRLB than its LoS counterpart. This is because each NLoS path provides three distinct parameters (AoA, AoD, delay), offering richer data for sensing compared to the two parameters available from a reciprocal LoS path.

Similarly, Fig. \ref{CRLBdelay} illustrates the CRLB versus delay error $\sigma_\tau$, revealing trends consistent with Fig. \ref{CRLBangle}. The CKM-based methods achieve significantly lower CRLBs than the geometry-based LoS method, which is attributed to their precise channel-to-location mapping via NLoS paths. Consistent with previous findings, the CKM-based NLoS approach outperforms both the CKM-based NLoS\_conv and LoS methods due to the richer channel knowledge it leverages, aligning with both the angle error and MSE analyses.

The simulation results demonstrate the potential of the proposed CKM-based sensing framework in ISAC systems. By leveraging multipath information and accurately predicting location-specific channel knowledge using an FCNN, the CKM-based method achieves significant improvements in sensing accuracy in NLoS scenarios compared to traditional geometry-based approaches. The CRLB analysis further confirms the robustness of the CKM-based method to angle and delay estimation errors. These findings highlight the effectiveness of the CKM framework for environment-aware sensing in complex wireless environments.

\section{Conclusion}
This paper introduces a CKM-enabled ISAC framework that achieves robust environment-aware sensing, even in challenging NLoS scenarios where traditional geometry-based methods fail. We demonstrate that the same CKM designed for environment-aware communications can be exploited to achieve sensing as well. Simulation results demonstrate significant performance gains over geometry-based approaches, with CRLB analysis confirming enhanced robustness. This approach marks a significant advancement for ISAC systems, enabling seamless integration of sensing and communication in complex environments using the same CKM.

\bibliographystyle{IEEEtran}
\bibliography{ref2}

\end{document}